\renewcommand{\vec}[1]{\mbox{\boldmath $#1$}}

\documentclass[preprint2]{pp7}

\usepackage{scalerel}
\newcommand\HII{H\protect\scaleto{$II$}{1.2ex}}
\newcommand\HI{H\protect\scaleto{$I$}{1.2ex}}
\newcommand\Malf{\mathcal{M}_{A}}
\usepackage[dvipsnames,table]{xcolor}

\newcommand{\frass}{FrASS}
\newcommand{\jai}{JAI}

\usepackage{enumitem}
\usepackage{placeins}
\usepackage{savesym}
\savesymbol{captionbox}
\usepackage{caption}
\restoresymbol{CAPTIONBOX}{captionbox}
\usepackage{url}
\def\stitle{Magnetic fields in star formation: from clouds to cores}                    
\def\sauthors{K. Pattle, L. Fissel, M. Tahani, T. Liu, E. Ntormousi}
\usepackage{times}
\usepackage{amsmath}

\usepackage{fancyhdr}

\fancypagestyle{fancy1st}{                      %Define style "fancy1st"
    \fancyhead[L]{}                             %Overwrite Left   Header
    \fancyhead[C]{}                             %Overwrite Center Header
    \fancyhead[R]{}                             %Overwrite Right  Header

    \fancyfoot[L]{Protostars and Planets VII}   %Overwrite Left   Footer
    \fancyfoot[C]{}                             %Overwrite Center Footer
    \fancyfoot[R]{}                             %Overwrite Right  Footer
}
\begin{document}

\title{\textbf{\LARGE Magnetic fields in star formation: from clouds to cores }}

\author {\textbf{\large Kate Pattle}}
\affil{\small\it Department of Physics and Astronomy, University College London, Gower Street, London WC1E 6BT, United Kingdom}
\affil{\small\it Centre for Astronomy, Department of Physics, National University of Ireland Galway, University Road, Galway H91 TK33, Ireland}
\author {\textbf{\large Laura Fissel}}
\affil{\small\it Department of Physics, Engineering Physics and Astronomy, Queen's University, Kingston, ON K7L 3N6, Canada}
\author {\textbf{\large Mehrnoosh Tahani}}
\affil{\small\it Dominion Radio Astrophysical Observatory, Herzberg Astronomy and Astrophysics Research Centre, National Research Council Canada, P. O. Box 248, Penticton, BC V2A 6J9, Canada }
\author {\textbf{\large Tie Liu}}
\affil{\small\it Shanghai Astronomical Observatory, Chinese Academy of Sciences, 80 Nandan Road, Shanghai 200030, People’s Republic of China }
\author {\textbf{\large Evangelia Ntormousi}}
\affil{\small\it Scuola Normale Superiore di Pisa, Piazza dei Cavalieri 7, 56126, Pisa (PI), Italy}

\begin{abstract}
\baselineskip = 11pt
\leftskip = 1.5cm 
\rightskip = 1.5cm
\parindent=1pc
{\small
In this chapter we review recent advances in understanding the roles that magnetic fields play throughout the star formation process, gained through observations and simulations of molecular clouds, the dense, star-forming phase of the magnetised, turbulent interstellar medium (ISM).
Recent results broadly support a picture in which the magnetic fields of molecular clouds transition from being gravitationally sub-critical and near equipartition with turbulence in low-density cloud envelopes, to being energetically sub-dominant in dense, gravitationally unstable star-forming cores.
Magnetic fields appear to play an important role in the formation of cloud substructure by setting preferred directions for large-scale gas flows in molecular clouds, and can direct the accretion of material onto star-forming filaments and hubs.
Low-mass star formation may proceed in environments close to magnetic criticality; high-mass star formation remains less well-understood, but may proceed in more supercritical environments.  The interaction between magnetic fields and (proto)stellar feedback may be particularly important in setting star formation efficiency.  We also review a range of widely-used techniques for quantifying the dynamic importance of magnetic fields, concluding that better-calibrated diagnostics are required in order to use the spectacular range of forthcoming observations and simulations to quantify our emerging understanding of how magnetic fields influence the outcome of the star formation process.
 \\~\\~\\~}
 %leave this in to get the correct vertical space after the abstract
\end{abstract}  

%\maketitle

\section{\textbf{INTRODUCTION}}
\label{sec:intro}

\begin{figure*}[!t]
    \centering
    \includegraphics[width=0.85\textwidth]{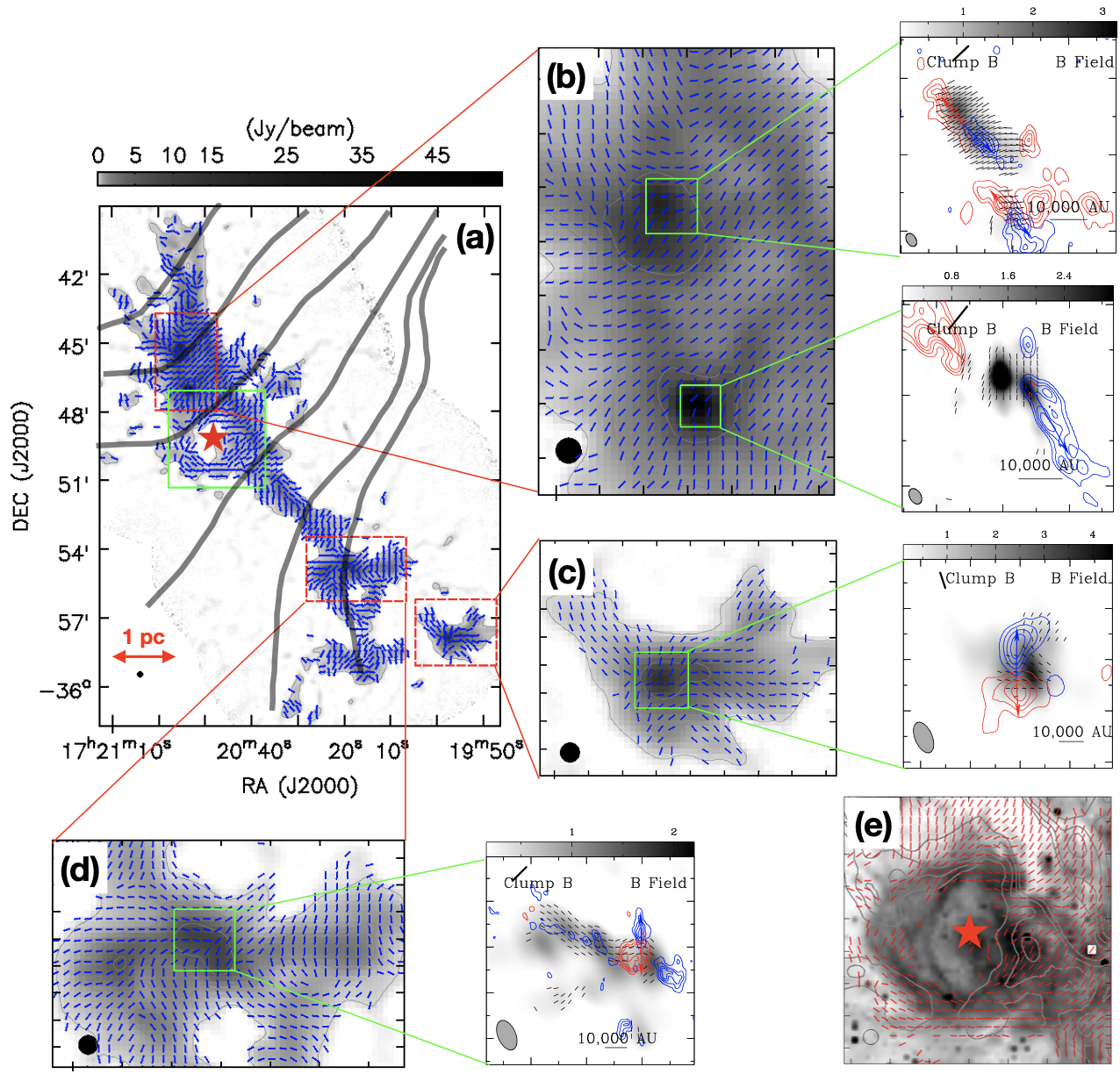}
    \caption{Magnetic fields inside the filamentary cloud NGC 6334. (a) The blue segments show magnetic field orientations obtained with JCMT/POL-2 at 850 $\mu m$ with a pixel size of 12$^{\prime\prime}$ \citep{arzoumanian2020}. The background image shows 850 $\mu m$ Stokes I emission.  The gray curves show the magnetic field lines inferred from SPARO polarization observations \citep{Lihuabai2015}. (b-d) The blue segments show magnetic field orientations obtained with JCMT/POL-2 at 850 $\mu m$ with a pixel size of 4$^{\prime\prime}$  \citep{arzoumanian2020}. The zoom-in aside panels show magnetic field orientations superposed on the Stokes I continuum emission at 345 GHz, observed by SMA at dense core scale \citep{Zhang2014}. The blue and red contours show the high-velocity CO outflow emission \citep{Zhang2014}. (e) Red segments show magnetic field orientations in the region outlined by the green box in panel a. The contours show 850 $\mu m$ Stokes I emission. The contour levels are 0.2, 0.4, 0.6 and 0.8 Jy~beam$^{-1}$. The red stars in panels a and e represent the H{\sc ii} region ``source II'' or ``D". The background image shows the Spitzer/IRAC 3.6 $\mu$m emission, from which the infrared bubble produced by the H{\sc ii} region is clearly seen. The magnetic fields are reshaped by the H{\sc ii} region and curved along the compressed shell. }
    \label{fig:NGC6334}
    \vspace{-0.5cm}
\end{figure*}

Star formation occurs within clouds of molecular hydrogen, the densest phase of the interstellar medium (ISM).  However, the turbulence, magnetization and gravity of the ISM turn star formation into a multi-scale process.  Molecular clouds are 
weakly ionized by UV photons and cosmic rays \citep{mckee1977} and, therefore, are coupled to the Galactic magnetic field \citep{mestel1956} at all but the very highest densities {in pre-star-formation gas} \citep{caselli1998}.
Throughout its volume, the ISM is therefore a magnetised turbulent fluid, in which stars form in the highest-density, gravitationally unstable regions of molecular clouds
\citep[e.g.,][]{benson1989}. Star formation is not the end-point of the process of ISM evolution; feedback effects from young stars and supernovae play a significant role in molecular cloud evolution and driving ISM turbulence \citep[e.g.,][]{krumholz2014}.  Fig.~\ref{fig:NGC6334} illustrates the ordered magnetic fields that pervade star-forming molecular clouds on all size scales. 
These magnetic fields can affect the evolution of star-forming regions in many ways: for example by altering the characteristics of turbulence \citep[e.g.,][]{brandenburg2013}, changing the characteristics of shocks \citep[e.g.,][]{inoue2009}, providing directionality to gas flows \citep[e.g.,][]{soler2013,Seifried2015}, providing pressure support against gravitational instability \citep[e.g.,][]{nakano1978}, removing angular momentum \citep[e.g.,][]{allen1993}, and transporting feedback \citep[e.g.,][]{offner2018} and cosmic rays \citep[e.g.,][]{shukurov2017} over large scales.

Stars form inefficiently; the Galactic star formation rate is orders of magnitude lower than if clouds were in a state of freefall collapse \citep[e.g.,][]{scalo1986}, and molecular clouds typically convert only a few percent of their mass into stars \citep[e.g.,][]{leisawitz1989}.  
Both magnetic fields and turbulence have been invoked as a means of supporting molecular clouds against gravitational collapse.  In the past, theories of the role of magnetic fields have tended toward the extremes: star formation either as a secular process mediated by ambipolar diffusion (ion-neutral drift) in a magnetically-dominated ISM \citep[e.g.,][]{shu1987}, or as the result of dynamic cloud evolution driven by supersonic and super-Alfv\'enic turbulence \citep[e.g.,][]{maclow2004}.  Such disparate models have developed in parallel in large part due to the significant difficulties in measuring interstellar magnetic field morphology and strength \citep[e.g.,][]{crutcher2012}, and in understanding and modelling the properties of magnetohydrodynamic (MHD) turbulence and so making testable predictions for magnetic field behavior \citep[e.g.,][]{KFreview19}.

However, a more complete observational and theoretical understanding 
has emerged over the last decade, thanks to major advances in instrumentation \citep[e.g.,][]{lamarre2010,friberg2016,cortes2016} and the frequent inclusion of magnetic fields into simulations of ISM physics \citep[e.g.,][]{VS11, Seifried2015, LiKlein2019}.  
The increasing community interest in 
combining our observational and theoretical knowledge is reflected in the direct comparison of theory and observations through production of synthetic observations of simulations \citep[e.g.,][]{soler2013,Seifried2019,king2018}.

In this chapter we present a  multi-scale review of magnetic fields in star-forming regions, ranging from
the formation of giant molecular cloud complexes ($\sim 100$\,pc) to dense cores forming individual
stellar systems ($\sim 0.01$\, pc), focusing in particular on these key questions:\\
\textbullet\ What are the three-dimensional (3D) morphologies of the magnetic fields of molecular clouds and their substructures, and do the observed field properties agree with different formation models of clouds, filaments and cores? \\
\textbullet\ Do magnetic fields direct gas accretion onto dense substructures, or are they distorted by gas motions, and how well coupled is the magnetic field to the gas in different density regimes? \\
\textbullet\ How does the energy balance between magnetic fields, turbulence and gravity change as a function of density and size-scale, and can variations in this balance lead to differences in cloud structure and star formation efficiency?

We first discuss the current state of instrumentation and key metrics and methods for measuring the strength and dynamic importance of magnetic fields in \S \ref{sec:methods}.  We review magnetic fields in molecular clouds in \S \ref{sec:clouds}, in dense, star-forming filaments in \S \ref{sec:filaments}, and in dense molecular cores in \S \ref{sec:cores}.  In \S \ref{sec:synthesis} we present a synthesis of these results, revisiting our key questions to discuss our current understanding of how magnetic fields affect the star formation process.
Finally, we discuss forthcoming observations and simulations over the next five years, and how these may address the outstanding issues in this rapidly evolving field.

\section{\textbf{MEASUREMENTS, METRICS \& METHODS}}
\label{sec:methods}

\FloatBarrier

\subsection{\textbf{Recent advances in instrumentation}}
\label{sec:instrumentation}

In the last decade there have been {major} advances in {polarimetric instrumentation,} particularly in the development of far-infrared {(FIR)} and (sub)millimeter cameras sensitive to polarized dust emission.  Perhaps the most important advance has been the $Planck$ satellite \citep{lamarre2010}, which produced all-sky 353\,GHz (850$\mu$m) dust polarization maps \citep{planck2014-a09}.  Significant advances have also been made {on the ground:} the James Clerk Maxwell Telescope (JCMT)'s POL-2 polarimeter \citep{friberg2016} on the SCUBA-2 camera \citep{holland2013} operates at 850$\mu$m and 450$\mu$m. The Atacama Pathfinder Experiment (APEX) telescope also offered the PolKa polarimeter at 870$\mu$m \citep{wiesemeyer2014}.  The Atacama Large Millimeter/submillimeter Array (ALMA) has developed polarization {capabilities \citep{cortes2016,hull2020a},} while the Submillimeter Array (SMA) has {an upgraded correlator \citep{primiani2016}.}

Stratospheric polarimetry is becoming increasingly important, particularly in the absence of any new space-based polarimeters in the intermediate term. The Stratospheric Observatory for Infrared Astronomy (SOFIA)'s HAWC+ camera \citep{harper2018} operates in five bands from 53$\mu$m to 214$\mu$m, while the {Balloon-borne Large-Aperture Submillimetre Telescope for Polarimetry (BLASTPol; (250$\mu$m, 350$\mu$m, 500$\mu$m)} \citep{galitzki2014b} and PILOT (214$\mu$m) \citep{foenard2018} telescopes have flown from different launch sites around the world.

{The new} large-detector-count cameras on single-dish instruments have made wide-area polarimetric surveys feasible. JCMT/POL-2 and SOFIA/HAWC+ have dedicated surveys of large areas of molecular clouds at resolutions of $\sim 10^{\prime\prime}$ in polarized light \citep[e.g., the JCMT BISTRO Survey;][]{wardthompson2017}. {Moreover, optical} and near-IR polarimeters have provided detailed large-scale maps of magnetic fields in low-density regions of molecular clouds, including SIRPOL on the InfraRed Survey Facility \citep[IRSF;][]{kandori2006}, Pico dos Dias \citep{Magalhaes1996}, the ARIES Imaging Polarimeter (AIMPOL) on the Sampurnanand telescope \citep{rautela2004}, and Mimir on the Perkins Telescope \citep{clemens2007,Clemens2020}.

\subsection{\textbf{Key ISM magnetic field tracers}}

\subsubsection{\textbf{Zeeman splitting of spectral lines}}
\label{sec:zeeman}

Line-of-sight magnetic field strengths can be directly measured through Zeeman splitting of spectral lines of paramagnetic species, observed either in absorption or emission\footnote{See, e.g., \citet{crutcher2019} for an introduction to the physics of the Zeeman effect.}.  In species with an unpaired electron, the line shifts induced by the Zeeman effect $\Delta\nu_{z}\propto \mu_{\textsc{b}}B$, where $\mu_{\textsc{b}}$ is the Bohr magneton.  The Zeeman effect has been detected in extended gas in H\textsc{i}, OH and CN.  H\textsc{i} in emission traces the cold neutral medium at hydrogen number densities $\sim 10^{0} - 10^{2}$\,cm$^{-3}$; OH emission and H\textsc{i} absorption trace a similar range of densities, $\sim 10^{2}-10^{4}$\,cm$^{-3}$, and CN traces densities $\sim 10^{5}-10^{6}$\,cm$^{-3}$.  The Zeeman effect can in principle give information on both the line-of-sight (LOS) and plane-of-sky (POS) components of the magnetic field ($B_{\textsc{los}}$ and $B_{\textsc{pos}}$ respectively); splitting due to the LOS component is seen in the Stokes $V$ {(circular polarization)} spectrum, with amplitude $\propto (\Delta\nu_{z}/\sigma_{v})B_{\textsc{los}}$, where $\sigma_{v}$ is the characteristic width of the spectral line, while splitting due to the POS component is seen in the Stokes $Q$ and $U$ {(linear polarization)} spectra, with amplitude $\propto (\Delta\nu_{z}/\sigma_{v})^{2}B_{\textsc{pos}}$.  (See, e.g., \citealt{tinbergen1996} for definitions of the Stokes parameters.)  Typically, $\Delta\nu_{z}\ll\sigma_{v}$ and so only the LOS component (and its direction) can be recovered.  Detecting the LOS Zeeman effect is itself very observationally intensive and requires Stokes $V$ instrumental polarization to be very well-characterised.

The Zeeman effect is more easily observed in polarized maser emission, which arises from compact (10--100\,au) objects with high brightness temperatures and densities \citep[e.g.][]{crutcher2019}.  Maser emission probes the small-scale physics of the later stages of high-mass star formation.  Key masing species include OH, associated with ultra-compact HII regions \citep[e.g.][]{caswell2011}; H$_2$O, tracing outflow shocks and protostellar discs \citep[e.g.][]{vlemmings2006}; and CH$_{3}$OH, tracing outflow shocks from high-mass star-forming regions (Class I), and the vicinities of massive protostars (Class II) \citep[e.g][]{cyganowski2009}.  The accuracy of magnetic field strength measurements in masing regions is being improved by modelling, including of non-Zeeman maser polarization \citep{lankhaar2019,dallolio2020}.

The Zeeman effect is a `gold standard' to which indirect measurements of {ISM} magnetic field strength are typically benchmarked \citep[e.g.,][]{heiles2009,poidevin2013}.  Despite this there are some caveats to Zeeman-derived magnetic field strengths: measurements are subject to line-of-sight reversals and beam integration effects \citep[e.g.,][]{poidevin2013}, and the significant time required to make the observations, particularly in higher-density gas traced by CN, mean that measurements at high densities are biased toward high-mass regions \citep{crutcher1999,falgarone2008}.  Only a handful of new {non-masing} Zeeman measurements have been published in the last decade {\citep{pillai2016,Thompson2019,Ching2022}}.

\subsubsection{\textbf{Faraday rotation}}
\label{sec:faraday}

When a linearly polarized electromagnetic (EM) wave passes through a magnetised region that contains free electrons (magnetized plasma), its plane of polarization rotates. This phenomenon is known as Faraday rotation and the amount of rotation can be obtained by:
%\begin{linenomath*}
\begin{equation}
\Delta \psi = \lambda^2 \big(0.812 \int{n_e \vec{B}\cdot \vec{dl}}\big) = \lambda^2 \rm{RM},
\end{equation}
%\end{linenomath*}
where $\Delta \psi$ is the amount of rotation (rad), $\lambda$ is the wavelength of the EM wave
(m), $n_e$ is the electron volume density of the magnetized region (cm$^{-3}$), $\vec{B}$ is the magnetic field strength ($\mu$G), and $\vec{dl}$ is the path length (pc). The quantity in brackets is the rotation measure (RM; rad\,m$^{-2}$).  Faraday rotation occurs because the ISM acts as a birefringent medium in the presence of magnetic fields and free electrons, resulting in different refractive indices for right- and left- circularly polarized EM waves. Faraday rotation provides information about the component of the magnetic field along the LOS. Interstellar synchrotron emission is a source for linearly polarized EM waves.

Traditionally, Faraday rotation of linearly polarized emission from pulsars and extragalactic compact sources was used to study galactic magnetic fields~\citep[e.g.,][]{Brownetal2007, VanEcketal2011} or the magnetic field of strongly ionized regions within the Galaxy~\citep{Harvey-Smithetal2011}. Various RM catalogs are available~\citep[e.g,][]{Schnitzeleretal2019, VanEcketal2021}; {currently, the most extensive is that of \cite{Tayloretal2009}, although being made at only two wavelengths, the uncertainty ranges in their derived RM values are relatively high.} 

Following the development of RM synthesis techniques \citep{Burn1966, BrentjensDeBruyn2005}, more information could be extracted about the ISM magnetic fields, including cold neutral \HI\ filaments \citep[e.g.,][using Low-Frequency Array, LOFAR, observations]{Zaroubietal2015, VanEcketal2019, Braccoetal2020a} or the foregrounds of \HII\ regions~\citep[][using the Parkes 64-m Radio Telescope as part of the Global Magneto-Ionic Medium Survey]{Thomsonetal2019}.  {It was thought that molecular clouds have $\sim$ zero contribution to the RM due to low abundance of free electrons.} However, \citet{Tahanietal2018} showed that a combination of stronger magnetic fields, the presence of free electrons in these regions due to cosmic rays, and higher densities resulting in higher electron densities even with lower ionization fractions, can result in an observable RM in these regions.  Even though UV fields can be strongly attenuated in dense molecular clouds, cosmic rays are an important source of ionization in these regions~\citep[e.g.,][]{Berginetal1999, EverettandZweibel2011, Padovanietal2018} and the ionization rates in denser regions can be higher than previously thought~\citep{Padovanietal2018}.

\citet{Tahanietal2018} developed a new technique using Faraday rotation from extragalactic sources and pulsars to determine the LOS magnetic field component in molecular clouds. This technique exploits an on-off approach to decouple the RM contribution by the cloud from the Galactic contribution {(everything along the LOS except the cloud).}  They then used extinction maps and a chemical evolution code to estimate the electron column densities and {so} the LOS magnetic field strength. They found that their obtained magnetic field directions were consistent with atomic Zeeman observations in the envelopes of molecular clouds.

\subsubsection{\textbf{Dust extinction/emission polarization}}
\label{sec:dust}

{Interstellar dust polarization \citep{hall1949, hiltner1949} in most ISM environments arises from grains aligned with their minor axis parallel to the magnetic field \citep{davis1951}, causing preferential polarization of dust-extincted optical and near-infrared (NIR) emission parallel to, and of far-IR and (sub)millimeter dust continuum emission perpendicular to, the POS magnetic field direction.  Radiative Alignment Torques (B-RATs; \citealt{dolginov1976, lazarian2007}) is the leading theory of grain alignment, in which paramagnetic grains are spun up by a non-isotropic radiation field to precess around the local magnetic field direction \citep{andersson2015}\footnote{Various alternative mechanisms are proposed in extremely high-density and/or strongly irradiated environments, including Mechanical Alignment Torques \citep{lazarian2007a, hoang2016}, k-RAT alignment \citep{lazarian2007, tazaki2017}, and dust self-scattering in protostellar discs \citep{kataoka2015}. {These mechanisms generally do not apply in the environments discussed in this review, {but may become particularly important in high-resolution observations of protostellar sources \citep{legouellec2020}.}}}}

{Dust polarization is a key tool 
because it allows plane-of-sky magnetic field direction to be traced over large areas and a wide range of densities relatively inexpensively.  Polarization fraction is at a maximum of $\sim 0.2$ in the low-density ISM \citep{planck2014-XIX}, and typically decreases significantly with increasing gas column density, to $< 0.01$ in starless cores \citep[e.g.,][]{jones2015}.  This `polarization hole' effect may be caused by some combination of loss of grain alignment at high visual extinction ($A_{V}$) \citep[e.g.,][]{whittet2008}, integration of complex field geometries within a telescope beam (`field tangling') \citep[e.g.,][]{hull2014}, and centrifugal destruction of dust grains by radiative torques in the immediate vicinity of protostars (`radiative torque disruption', RATD; \citealt{hoang2019}).  Interferometric observations of protostellar systems show that grains can remain aligned (e.g., \citealt{kwon2019}), but in these sources there is an internal source of photons to drive grain alignment.  Observations of the starless core FeST 1-457 have suggested that grains are unaligned beyond $A_{V}\sim 20-30$ mag \citep{alves2014,jones2015}; however, externally-illuminated starless cores may retain some degree of grain alignment to high $A_{V}$ \citep{pattle2019}.  \citet{hoang2021} proposed an analytical model in which the maximum $A_{V}$ at which grains remain aligned varies as a function of incident radiation anisotropy, gas density and grain size, among other parameters, with larger grains remaining aligned to significantly higher $A_{V}$.}

\subsubsection{\textbf{Goldreich-Kylafis effect}}
\label{sec:gk}

Emission line polarization can arise from the Goldreich-Kylafis (GK) effect \citep{goldreich1981}, in which molecular line emission may
be linearly polarized either parallel or perpendicular to the plane-of-sky magnetic field.  {The GK effect, which can provide a velocity-resolved probe of magnetic field morphology, has been observed in outflows \citep[e.g.][]{girart1999,ching2017}, and in the high-mass star-forming region NGC6334I(N) using ALMA \citep{cortes2021a}.}
{However, the uncertainty on polarization direction complicates interpretation.  ALMA's ability to measure both line and continuum polarization in a single spectral set-up will be key to understanding under what conditions the GK effect produces parallel or perpendicular alignment {\citep[e.g.][]{cortes2021a}}.  {A further complication is conversion of linear to circular line polarization by anisotropic resonant scattering by foreground material \citep{houde2013,chamma2018}.}}

\subsubsection{\textbf{Velocity gradients}}
\label{sec:vgt}

{The Velocity Gradient Technique (VGT; \citealt{gonzalez-casanova2017}), is a new method for inferring magnetic field morphologies, primarily in the low-density ISM.  VGT makes use of the elongation of turbulent eddies in the ISM along the local magnetic field direction, positing that fast turbulent magnetic reconnection across these eddies results in turbulent fluid motions being preferentially perpendicular to the magnetic field.  {Tests}
against simulations suggest that an optically thin gas tracer can be used to trace magnetic fields in regions with supersonic, trans- and sub-Alfv\'{e}nic turbulence and without strong gravitational collapse \citep{hsieh2019}, and the method has reproduced the large-scale magnetic field morphology of nearby GMCs with reasonable accuracy \citep{Hu2019}.} 

{VGT assumes that the velocity gradients seen in {thin} channel maps are associated with turbulent rather than density structures \citep{gonzalez-casanova2017}.  However, H\textsc{i} structures in the diffuse ISM have been shown to be associated with density enhancements \citep{clark2019}.
A striking alignment between velocity gradients and magnetic field directions is seen in the low-density, non-self-gravitating ISM, but debate continues over the physical origin of this effect \citep[e.g.,][]{kalberla2020}. 
A major strength of VGT is that it is velocity-resolved, probing the magnetic field structures of multiple velocity components along a single LOS \citep[e.g.,][]{Hu2019}.  Moreover, where the assumptions of VGT break down may be a good probe of the transition of the ISM to the gravity-dominated regime \citep[cf.][]{hu2020,Hu2021Serp}, and VGT could be used to make predictions for where this transition occurs.}

\subsection{\textbf{Metrics and methods}}

\subsubsection{\textbf{Key metrics}}
\label{sec:metrics}

{We here outline the key metrics by which the relative importance of magnetic fields in the ISM is parameterized, the values of which may change with size and density scale.}

{\textbf{Energy balance}  
Magnetic energy is given by }
%\begin{linenomath*}
\begin{equation}
    E_{B} = \frac{B^{2}V}{2\mu_{0}}\,\,({\rm SI}) = \frac{B^{2}V}{8\pi}\,\,({\rm cgs}),
\end{equation}
%\end{linenomath*}
{where $V$ is volume, and can be compared to the other energy terms, typically gravitational potential energy, thermal or non-thermal kinetic energy and external pressure energy. $E_{B}$ is typically subject to large uncertainties.} 

\textbf{Mass-to-flux ratio}  The critical mass-to-flux ratio is
%\begin{linenomath*}
\begin{equation}
    \left(\frac{M}{\Phi}\right)_{crit}=\frac{1}{2\pi\sqrt{G}}\,\,{\rm (cgs)},
\end{equation}
%\end{linenomath*}
{\citep{nakano1978}.  The mass-to-flux ratio is then given in units of the critical value by}
%\begin{linenomath*}
\begin{equation}
    \mu_{\Phi} = \frac{(M/\Phi)}{(M/\Phi)_{crit}} = 2\pi\sqrt{G}\mu m_{\textsc{h}}\left(\frac{N}{B}\right)\,\,{\rm (cgs)},
\end{equation}
%\end{linenomath*}
{where $N$ is column density, $B$ is magnetic field strength, and $\mu$ is mean molecular weight.  A value $\mu_{\Phi} > 1$ indicates that the region is \emph{magnetically supercritical}, i.e. the magnetic field cannot prevent gravitational collapse, while $\mu_{\Phi} < 1$ indicates that the region is \emph{magnetically subcritical}, i.e. magnetically supported.  Geometric corrections to $\mu_{\Phi}$ can be significant \citep{crutcher2004}.}

{\textbf{Alfv\'en Mach number}  The Alfv\'en velocity,}
%\begin{linenomath*}
\begin{equation}
    v_{A} = \frac{B}{\sqrt{\mu_{0}\rho}}\,\,({\rm SI}) = \frac{B}{\sqrt{4\pi\rho}}\,\,({\rm cgs}),
\end{equation}
%\end{linenomath*}
{where $\rho$ is gas density and $\mu_{0}$ is the permeability of free space, is the group velocity of transverse oscillations of matter and magnetic field lines, for which magnetic tension is the restoring force. This 
must be measured across magnetic field lines; $v_{A}$ along a field line is infinite.  The relative importance of magnetism and non-thermal gas motions is parameterised by the Alfv\'en Mach number,}
%\begin{linenomath*}
\begin{equation}
    \mathcal{M}_{A} = \frac{\sigma_{v,\textsc{nt}}}{v_{A}} \propto\left(\frac{E_{K,\textsc{nt}}}{E_{B}}\right)^{0.5},
\end{equation}
%\end{linenomath*}
where $\sigma_{v,\textsc{nt}}$ is the non-thermal velocity dispersion, $E_{B}$ is magnetic energy and $E_{K,\textsc{nt}}$ is non-thermal kinetic energy.  {If turbulence is isotropic then $\sigma_{v,\textsc{nt}}$ should be 3D, i.e. $\sqrt{3}\times$ its measured LOS value \citep{crutcher1999}.  However, turbulence will be anisotropic in the presence of a strong mean magnetic field.} $\mathcal{M}_{A} < 1$ (sub-Alfv\'enic) indicates that magnetic fields direct gas motions; $\mathcal{M}_{A} > 1$ (super-Alfv\'enic) indicates the converse. $\mathcal{M}_{A}$ is analogous to sonic Mach number, $\mathcal{M}=\sigma_{v,\textsc{nt}}/c_{s}$, where $c_{s}$ is sound speed.

{\textbf{Plasma beta}  The thermal-to-magnetic pressure ratio is}
%\begin{linenomath*}
\begin{equation}
    \beta = \frac{nk_{\textsc{b}}T}{B^{2}/2\mu_{0}}\,\,({\rm SI}) = \frac{nk_{\textsc{b}}T}{B^{2}/8\pi}\,\,({\rm cgs}) = \frac{E_{K,\textsc{t}}}{E_{B}},
\end{equation}
%\end{linenomath*}
{where $T$ is temperature, $n$ is number density, and $E_{K,\textsc{t}}$ is thermal kinetic energy.  A value of $\beta \ll 1$ indicates a magnetically-dominated system; $\beta \gg 1$ indicates a thermally-dominated system.}

{\textbf{Jeans Mass}  The classic measure of stability of an isothermal gas sphere is the Jeans mass \citep{jeans1928},}
%\begin{linenomath*}
\begin{equation}
    M_{J} = \frac{4\pi}{3}\frac{c_{s}^{3}}{G^{3/2}\rho^{1/2}}.
\end{equation}
%\end{linenomath*}
{If non-thermal motions are taken to represent an effectively hydrostatic pressure against collapse (the microturbulent assumption, \citealt{chandrasekhar1951}), then $c_{s}$ can be replaced by $\sigma_{v,\textsc{nt}}$ or $(c_{s}^{2} + \sigma^{2}_{v,\textsc{nt}})^{0.5}$.  A structure with a mass significantly greater than its Jeans mass is suggestive of significant magnetic support (e.g., \citealt{sanhueza2019}).}

{\textbf{Virial balance} The overall energetic balance of a \emph{small-scale} cloud structure can be estimated in terms of its virial balance; however, this makes the assumption, invalid except on scales smaller than the thermal Jeans wavelength, $\lambda_J = c_{s}(\pi/G\rho)^{0.5}$, that the structure is evolving quasistatically, with turbulent motions providing mean support against gravity \citep{maclow2004}.}

{\textbf{Freefall time} An ISM structure in a state of unimpeded gravitational collapse will collapse on its freefall timescale,}
%\begin{linenomath*}
\begin{equation}
    t_{ff} = \left(\frac{3\pi}{32G\rho}\right)^{0.5}.
    \label{eq:t_ff}
\end{equation}
%\end{linenomath*}

{\textbf{Ambipolar diffusion timescale} A structure evolving quasistatically to instability in a strong magnetic field will have a lifetime set by the ambipolar diffusion timescale -- the characteristic timescale of magnetic flux loss as neutral species drift past ions \citep[e.g.,][]{heitsch2003}:
%\begin{linenomath*}
\begin{equation}
    t_{AD} = \frac{L^{2}}{\lambda_{AD}},
    \label{eq:t_ad}
\end{equation}
%\end{linenomath*}
where $L$ is the characteristic size of the structure in question and $\lambda_{AD}$ is the ambipolar diffusivity,
%\begin{linenomath*}
\begin{equation}
    \lambda_{AD} = \frac{\mu_{i}+\mu_{n}}{4\pi\langle\sigma v\rangle\mu_{i}\mu_{n}m_{\textsc{h}}x_{i}}\left(\frac{B}{n_{n}}\right)^{2}\,\,{\rm (cgs)},
    \label{eq:ambipolar_diff}
\end{equation}
%\end{linenomath*}
where $\mu_{i}$ and $\mu_{n}$ are the mean molecular weight of ions and neutrals respectively, $n_{n}$ is the number density of neutrals, $x_{i}$ is ionization fraction, and $\langle\sigma v\rangle$ is the rate coefficient for elastic collisions ($\langle\sigma v\rangle = 1.5\times 10^{-9}\,$cm$^{3}$s$^{-1}$; \citealt{draine1983}).  Various formulations of $t_{AD}$ exist for the dense, cosmic-ray-ionized ISM, many of which assume ionization-recombination balance, and so that
%\begin{linenomath*}
\begin{equation}
    x_{i}\propto \left(\frac{n_{n}}{\zeta}\right)^{-0.5}\approx 1.2\times10^{-5}n_{n}^{-0.5}\,\,({\rm cgs}),
\end{equation}
%\end{linenomath*}
where $\zeta$ is cosmic ray ionization rate \citep{elmegreen1979,umebayashi1980}.  A recent formulation of $t_{AD}$ \citep{heitsch2014} is
%\begin{linenomath*}
\begin{equation}
    t_{AD} = 1.38\times10^{3}\left(\frac{L}{1\,{\rm pc}} \right)^{2}\left(\frac{n}{300\,{\rm cm}^{-3}}\right)^{1.5}\left(\frac{B}{5\,\mu{\rm G}}\right)^{-2}\,\,{\rm Myr}.
    \label{eq:t_ad_emp}
\end{equation}
%\end{linenomath*}}

{\textbf{Magnetic field-density relation}
The relationship between $B$ and $n$ is typically parameterised as}
%\begin{linenomath*}
\begin{equation}
    B\propto n^{\kappa}.
    \label{eq:B_vs_n}
\end{equation}
%\end{linenomath*}
{
Collapse of a spherical cloud with flux-freezing should produce $\kappa\approx 2/3$ \citep{mestel1966}, while ambipolar diffusion models predict $0<\kappa<0.5$, evolving from $\kappa\sim 0$ initially (indicating collapse along field lines) to $\kappa\sim 0.5$ in the later stages of collapse (indicating collapse across field lines) \citep[e.g.,][]{mouschovias1999}.  
$\kappa > 0$ indicates that magnetic flux is increasing with density \citep[e.g.,][]{tritsis2015}, and so that the field is being compressed, typically by gravity, although stellar feedback in H\textsc{ii} regions could also cause such compression.  A widely used form of eq. \ref{eq:B_vs_n}, following \citet{crutcher2010}, is}
%\begin{linenomath*}
\begin{equation}
    B = 
    \begin{cases}
    B_{0} & (n < n_{0}) \\
    B_{0}\left(\dfrac{n}{n_{0}}\right)^{\kappa} & (n > n_{0})
    \end{cases}.
    \label{eq:B_vs_n_emp}
\end{equation}
%\end{linenomath*}

\subsubsection{\textbf{The Davis-Chandrasekhar-Fermi (DCF) method}}
\label{sec:dcf}

\begin{figure*}[!t]
    \centering
    \includegraphics[width=\textwidth]{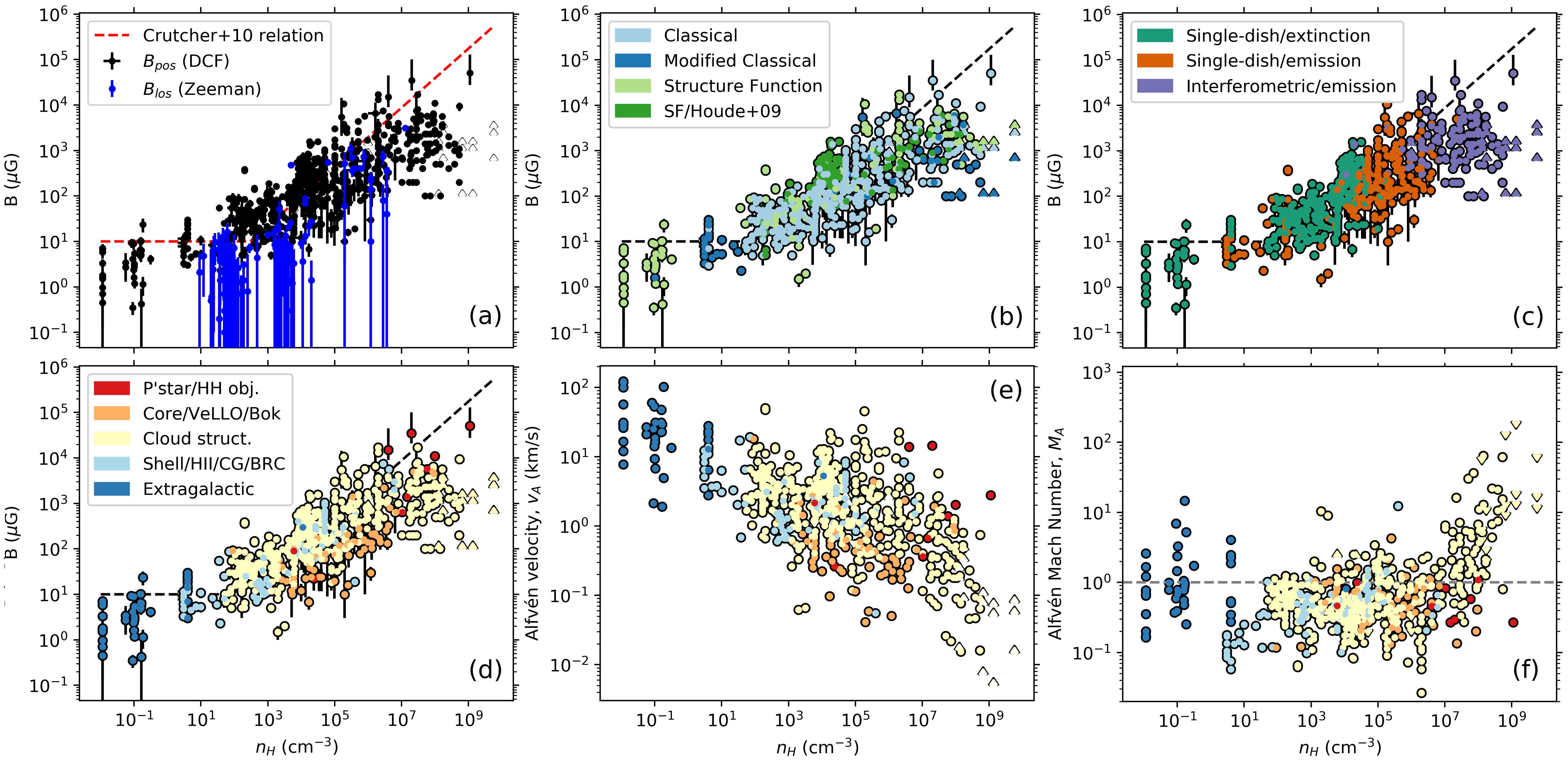}
    \caption{{(a)}: magnetic field strength as a function of hydrogen number density: black points show DCF measurements; blue show Zeeman measurements.  {Arrows show upper/lower-limit measurements.}  Red dashed line shows the \citet{crutcher2010} relation.  {(b)}: as {(a)}, but with DCF measurements only.  Data points are color-coded by DCF variant.  {(c)}: as {(b)}, color-coded by measurement type. {(d)}: as {(b)}, color-coded by object type. {(e)}: Alfv\'en velocity of the DCF measurements, color-coded by object type.  {(f)}: Alfv\'en Mach number of the DCF measurements, color-coded by object type.  Dashed line marks $\mathcal{M}_{A} = 1$.  References: {\scriptsize{\citet{alina2020}, \citet{alves2008,alves2011}, \citet{andersson2005,andersson2006}, \citet{Anez2020}, \citet{arzoumanian2020}, \citet{attard2009}, \citet{beltran2019}, \citet{bertrang2014}, \citet{beuther2010,beuther2018}, \citet{cashman2014}, \citet{chakraborty2016}, \citet{chapman2011}, \textit{H.-R. Chen et al.} (\citeyear{chenHR2012}), \textit{Z. Chen et al.} (\citeyear{chenZ2012,chenZ2017}), \citet{ching2017}, \citet{choudhury2019}, \citet{Chuss2019}, \citet{cortes2006,cortes2010,cortes2016,cortes2019,cortes2021}, \citet{coude2019}, \citet{crutcher2004}, \citet{curran2004,curran2007}, \citet{Dallolia2019}, \citet{das2016}, \citet{Devaraj2021}, \citet{dewangan2015,Dewangan2018}, \citet{eswaraiah2013,eswaraiah2017,eswaraiah2019,eswaraiah2020,eswaraiah2021}, \citet{franco2015}, \citet{frau2014}, \citet{girart2006}, \citet{henning2001}, \citet{Heyer2008}, \citet{hildebrand2009}, \citet{hilyblant2007}, \citet{houde2009,houde2016}, \citet{hoq2017}, \citet{hull2017}, \citet{joubaud2019}, \citet{juarez2017}, \citet{kandori2017a,kandori2020_CrA-E,kandori2020_bhr71,kandori2020_B335,kandori2020_l1774,kandori2020}, \citet{karoly2020}, \citet{kim2016}, \citet{kirby2009}, \citet{kirk2006}, \citet{Konyves2021}, \citet{kusune2015,kusune2016}, \textit{J. Kwon et al.} (\citeyear{kwon2010,kwon2011,kwon2016,kwon2018}), \textit{W. Kwon et al.} (\citeyear{kwon2019}), \citet{lada2004}, \citet{lai2001,lai2002}, \citet{lee2014,lee2018}, \citet{li2011,Lihuabai2015}, \textit{J. Liu et al.} (\citeyear{liu2019,LiuJH2020}), \textit{T. Liu et al.} (\citeyear{liu2018,liu2018a}), \citet{lobogomez2015}, \citet{mao2008}, \citet{marchwinski2012}, \citet{matthews2002,matthews2005}, \citet{neha2016,neha2018}, \citet{ngoc2021}, \citet{Palau2021}, \citet{panopoulou2016}, \citet{pattle2017a,pattle2018,pattle2021}, \citet{pereyra2007}, \citet{pillai2015,pillai2016}, \citet{Planck_XXXV2016}, \citet{poidevin2006,poidevin2013}, \citet{qiu2013,qiu2014}, \citet{rao2009}, \citet{rathborne2009}, \citet{redaelli2019}, \citet{rodrigues2007}, \citet{sadavoy2018}, \citet{Santosetal2014,santos2016}, \citet{sharma2020}, \citet{soam2015,soam2015a,soam2017,soam2017a,soam2018,soam2018a,Soam2019,soam2019a,soam2021}, \citet{soler2018}, \citet{stephens2013}, \citet{sugitani2010,sugitani2011,Sugitani2019}, \citet{tamaoki2019}, \citet{tang2009,Tang2019}, \citet{tsuboi2021}, \citet{vallee2003,vallee2005,vallee2006,vallee2007,vallee2007a,vallee2007b}, \citet{wang2019,wang2020b}, \citet{wisniewski2007}, \citet{wolf2003}, \citet{wright2014}, \citet{zielinski2021}.}}}
    \label{fig:compilation}
    \vspace{-0.5cm}
\end{figure*}

{The Davis-Chandrasekhar-Fermi (DCF) method  \citep{davis1951a, chandrasekhar1953a} is a means of estimating $B_{\textsc{pos}}$ by taking dispersion in polarization angle from dust emission or extinction measurements to indicate distortion of the magnetic field by non-thermal gas motions, and so to be a measure of $\mathcal{M}_{A}$. 
For a given turbulent velocity dispersion and gas density, magnetic field strength can then be inferred.  The DCF equation in its original form is}
%\begin{linenomath*}
\begin{equation}
    B_{\textsc{pos}} = \sqrt{\mu_{0}\rho}\frac{\sigma_{v,\textsc{nt}}}{\sigma_{\theta}}\,\,({\rm SI}) =  \sqrt{4\pi\rho}\frac{\sigma_{v,\textsc{nt}}}{\sigma_{\theta}}\,\,({\rm cgs})
    \label{eq:dcf}
\end{equation}
%\end{linenomath*}
{where $\rho$ is gas density, $\sigma_{v,\textsc{nt}}$ is non-thermal linewidth in a gas species taken to trace the same material as the dust polarization observations, and $\sigma_{\theta}$ is dispersion in polarization position angle.  DCF %analysis
{makes} several assumptions, most notably that turbulence is sub-Alfv\'enic, but also that the underlying magnetic field geometry is linear, and that $\sigma_{v,\textsc{nt}}$ traces 
turbulent motions.  Nonetheless, it
provides an estimation of magnetic field strength from dust {polarization},
and so is widely used despite long-standing theoretical concerns \citep[e.g.,][]{zweibel1990,myers1991,houde2009}.  DCF measures an average $B_{\textsc{pos}}$ in the area over which $\sigma_{\theta}$ is measured; however, recent wide-area high-resolution polarimetric mapping of  molecular clouds has led to resolved DCF being used to map $B_{\textsc{pos}}$ variation across clouds \citep{Guerra2021, hwang2021}.}

{The original DCF method {likely overestimates}
$B_{\textsc{pos}}$ due {to} 
integration of ordered structure on scales smaller than the telescope beam, and from multiple turbulent cells within the beam {and}
along the LOS \citep[e.g.,][]{ostriker2001,houde2009}.  We outline the methods of accounting for this here; {see}
\citet{pattlefissel2019} for a detailed comparison.}

{`Classical' DCF modifies eq.~\ref{eq:dcf} by a factor $0<Q\leq 1$, generally $Q=0.5$, such that $(1/\sigma_{\theta})\to (Q/\sigma_{\theta})$ to account for integration effects \citep{ostriker2001,heitsch2001,padoan2001}.
\citet{cho2016} proposed the ratio of velocity centroid dispersion to linewidth as an estimator of the number of turbulent cells along the LOS.  Further modifications can be made to account for large-scale magnetic field structure when estimating $\sigma_{\theta}$ \citep{pillai2015,pattle2017a}.  Classical DCF is often restricted to regions where $\sigma_{\theta}<25^{\circ}$ \citep{heitsch2001}.}

{Alternatively, $\sigma_{v,\textsc{nt}}/\sigma_{\theta}$ in eq.~\ref{eq:dcf} can be replaced with the ratio of energies in the turbulent and ordered field components, such that $1/\sigma_{\theta}\to(\langle B_{t}^{2}\rangle/\langle B_{o}^{2}\rangle)^{-0.5}$.  This ratio is determined from the structure function of the dispersion in polarization angles \citep{falceta-goncalves2008}; \citet{hildebrand2009} proposed a means of accounting for large-scale field structure, and \citet{houde2009} further expanded the method to account for sub-beam and LOS effects.  \citet{lazarian2020} have recently proposed a variation 
using structure functions to also measure $\sigma_{v,\textsc{nt}}$.}

{The proliferation of DCF measurements in recent years has led to renewed interest in testing DCF variants against simulations.}  \citet{skalidis2021} have proposed an alternative DCF equation,
taking non-Alfv\'enic (compressible) modes to dominate Alfv\'enic (incompressible) modes,
%\begin{linenomath*}
\begin{equation}
    B_{\textsc{pos}} = \sqrt{2\pi\rho}\frac{\sigma_{v,\textsc{nt}}}{\sqrt{\sigma_{\theta}}}\,\,({\rm cgs}),
\end{equation}
%\end{linenomath*}
{tested against a range of ideal-MHD simulations by \citep{skalidis2021a}.  However, \citet{li2021} argue that while the compressible modes are significant, compressions and rarefactions largely cancel one another out.}
{\citet{li2021} present a derivation of, and through comparison with simulations advocate, replacement of $\sigma_{\theta}\to\tan\sigma_{\theta}$ in eq.~\ref{eq:dcf} (cf. \citealt{heitsch2001}; \citealt{falceta-goncalves2008}), thereby removing the small-angle restrictions on classical DCF.}

{\citet{liu2021} applied DCF to simulations, finding that}
$B_{\textsc{pos}}$ is accurately recovered in strong-field cases, but may be significantly overestimated in super-Alfv\'enic environments.  They propose an environment-dependent {$Q$ range}, 
and find that structure functions can
{characterize} ordered field structure and sub-beam integration effects, but that the \citet{cho2016} method may best estimate
{LOS} turbulent cells, 
{with DCF unreliable} on size scales $< 0.1$\,pc unless LOS integration is accounted for.
These results 
encapsulate the challenge of DCF: how its applicability can be judged without an independent measure of $\mathcal{M}_{A}$.

\subsubsection{\textbf{Compilation of literature DCF field strengths}}
\label{sec:dcf_compilation}

{Over the last few years, hundreds of DCF measurements have been published.  We have attempted to compile every DCF measurement published since the \citet{ostriker2001}, \citet{padoan2001} and \citet{heitsch2001} papers with clearly identifiable density and non-thermal velocity dispersion values\footnote{We investigated every paper published between 2001 and May 2021 citing \citet{chandrasekhar1953a}, and have made a best-efforts attempt to identify DCF studies citing \citet{chandrasekhar1953} in error.}.  These are shown in Fig.~\ref{fig:compilation}a, alongside the \citet{crutcher2010} Zeeman measurements.  There is a striking correlation between the two data sets; the DCF $B_{\textsc{pos}}$ estimates, while typically larger than the Zeeman $B_{\textsc{los}}$ measurements at a given density, are {largely} compatible with the $B\propto n^{0.65}$ relationship up to densities $\sim 10^{7}$\,cm$^{-3}$, and then become more broadly distributed.}

{We classified each measurement as `classical' (using eq.~\ref{eq:dcf} modified by a factor $0<Q\leq1$), `modified classical' (classical DCF modified following \citealt{heitsch2001, pillai2015, pattle2017a, cho2016}), `structure function' (following \citealt{falceta-goncalves2008} or \citealt{hildebrand2009}) or `SF/Houde+09' (following \citealt{houde2009}), as shown in Fig.~\ref{fig:compilation}b.  We classified each measurement as arising from either extinction, single-dish emission, or interferometric emission polarimetry, as shown in Fig.~\ref{fig:compilation}c.  We identified five categories of object: (1) protostar, jet or Herbig Haro object, (2) isolated starless or protostellar core, VeLLO or Bok globule, (3) `cloud structure', any GMC or structure within a GMC, including filaments, clumps and massive dense cores, (4) distinct structures under stellar feedback: shells, H\textsc{ii} regions, cometary globules and bright-rimmed clouds, and (5) extragalactic structures, as shown in Figs.~\ref{fig:compilation}d-f.  }

{
For DCF measurements the $n$ and $B_{\textsc{pos}}$ axes are not independent (cf. eq.~\ref{eq:dcf}), and so it is unsurprising that the results in Fig.~\ref{fig:compilation}a-d show a strong correlation.  The fundamental quantity being measured in DCF analysis is $\mathcal{M}_{A}$, and so we have attempted to recover this quantity for the measurements in our sample\footnote{$\mathcal{M_{A}}$ is defined for motion across field lines, while $\sigma_{\theta}$ traces the projection of magnetic field variations on the POS, and $\sigma_{v,\textsc{nt}}$ is measured along the LOS.  There is some inconsistency in the literature over whether the 1D or 3D velocity dispersion is appropriate in eq.~\ref{eq:dcf}; 
we use values as supplied in each paper.  This ambiguity has largely been subsumed in the wider uncertainties on DCF, but makes definitively identifying the Alfv\'enic state of measurements of $\mathcal{M}_{A}$ in the range $1/\sqrt{3} < \mathcal{M}_{A} < \sqrt{3}$ difficult.}.  We calculated $v_{A} = B/\sqrt{4\pi\rho}$, assuming $\rho = \mu m_{\textsc{h}}n_{\textsc{h}}$, and taking a mean particle weight $\mu=2.8$ for molecular gas and 1.4 for atomic gas,
as shown in Fig.~\ref{fig:compilation}e {(calculated from $B_{\textsc{pos}}$ only)}.  
}
{We then calculated $\mathcal{M}_{A} = \sigma_{v,\textsc{nt}}/v_{A} $ $ = \sigma_{\theta}/Q \,\,{\rm (classical)} = (\langle B_{t}^{2}\rangle/\langle B_{o}^{2}\rangle)^{0.5} \,\,{\rm (s.f.)}$,
as shown in Fig.~\ref{fig:compilation}f\footnote{In the few cases where we could not determine whether velocity dispersion values were given as Gaussian widths or FWHMs, we assumed the value was a Gaussian width.  {We place no limits on allowed values of $Q$ or $\sigma_{\theta}$, using the data as supplied in the original publications.}}.  $\mathcal{M}_{A}$ is broadly flat below $n_{\textsc{h}} \sim 10^7$\,cm$^{-3}$, albeit with scatter in the range 0.1--10.
The maximum $\mathcal{M}_{A}$ increases significantly at high densities.  The mean value of $\mathcal{M}_{A}$ below $10^7$\,cm$^{-3}$ is 0.74, and the median is 0.52 {(again calculated from $B_{\textsc pos}$ only)}, suggesting that turbulence is typically slightly sub-Alfv\'enic.  However, DCF assumes sub-Alfv\'enic turbulence and so it is {unsurprising} that we generally recover $\mathcal{M}_{A} < 1$.}

We discuss this {compilation} further in \S \ref{sec:dcf_discussion}.  {As} the analysis which we can perform in this chapter is very limited, we {have} made this data set available as a {resource}\footnote{See supplementary material at \url{http://ppvii.org/}}.  {We draw attention to a recent analysis of a compilation of emission DCF measurements by \citet{liu2021a}.}

\subsubsection{\textbf{The Histogram of Relative Orientations (HRO)}}
\label{sec:hro}

{{The HRO is commonly used in numerical and observational data to}}
{measure the alignment between density 
or column density structures and the local magnetic field \citep{soler2013}. The method calculates the angle $\phi$ between the local magnetic field and density gradient and its distribution in different density or column density bins. The scale-dependent behavior of the angle is expressed by an alignment parameter, which is positive (negative) when the magnetic field is predominantly parallel (perpendicular) to the density structures at a given bin. {Two commonly used parameters are the HRO shape parameter $
\xi = (A_{c} - A_{e})/(A_{c} + A_{e})$, where $A_{c}$ and $A_{e}$ are the number of measurements with $|\phi| \leq 22.5^{\circ}$ and $|\phi| \geq 67.5^{\circ}$ respectively, and the projected Rayleigh statistic $Z_{x} = \sum_i^n cos(2\phi_i)/\sqrt{n/2}$ \citep{jow2018}.}  The HRO can {thus} condense the statistical behavior of MHD turbulence {into} a single parameter, {making} %. This property has made 
the method particularly attractive for characterizing molecular clouds.

\begin{figure*}[!t]
    \centering
    \includegraphics[width=0.85\textwidth]{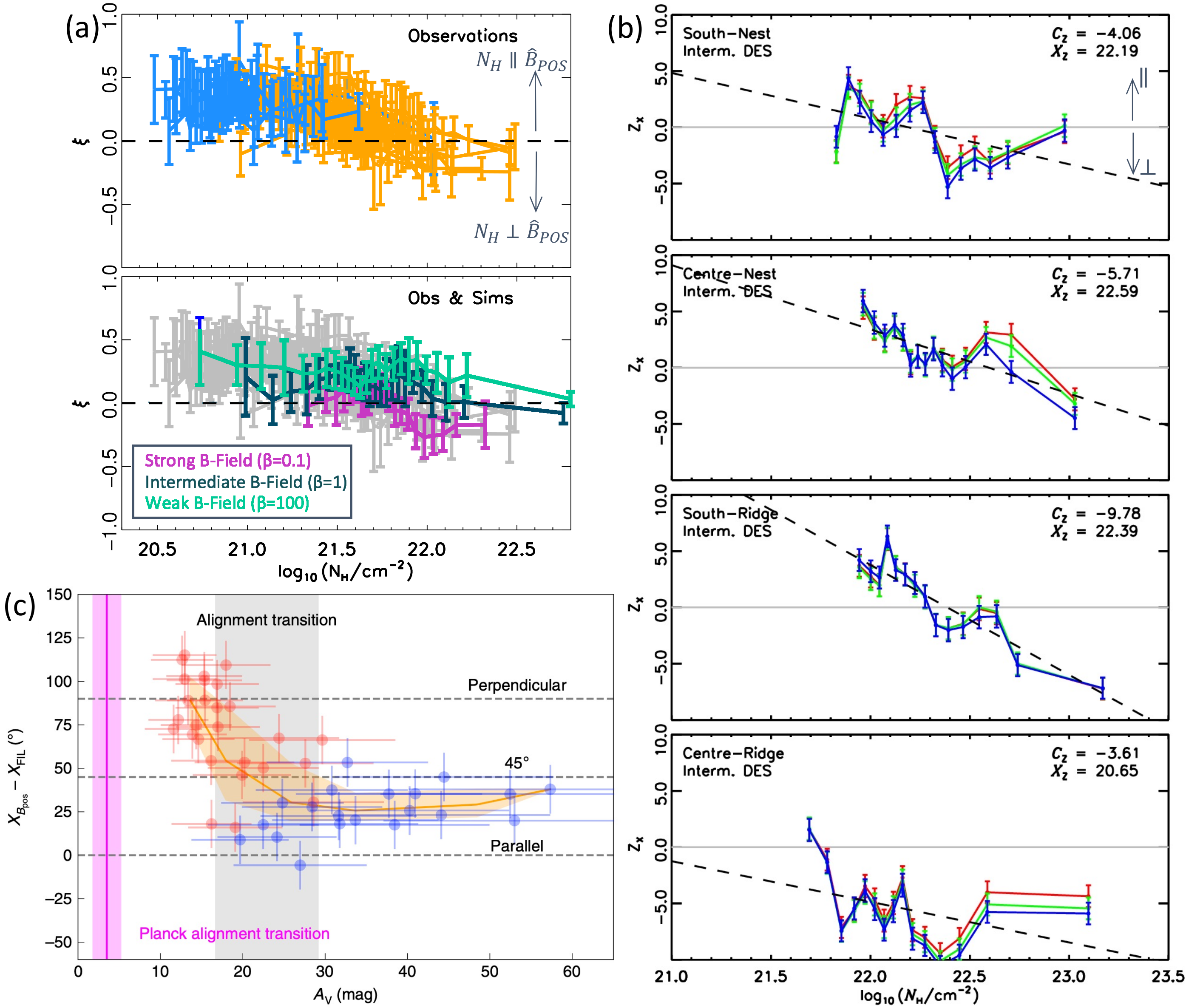}
    \caption{Observations of the relative alignment of cloud column density sub-structure with the magnetic field inferred from dust polarization maps. (a) Comparison of the HRO shape parameter $\xi$ for {\em Planck} 10$^{\prime}$\,FWHM 353\,GHz polarization observations of low column density clouds (blue) and a compilation of higher column density star-forming molecular clouds (orange), {\em bottom panel:} {\em Planck} $\xi$ vs.~$\log(N_{\mathrm H})$ observations (gray), compared to the $\xi$ vs.~$\log(N_{\mathrm H})$ found in simulations from \cite{soler2013} (adapted from \citealt{Planck_XXXV2016}; reproduced with permission \copyright\ ESO); {(b)} Relative alignment between {\em Herschel}-derived column density maps and BLASTPol 3-color polarization $\hat{B}_{\textsc{pos}}$ maps (red: 500\,$\mu$m, green 350\,$\mu$m, blue 350\,$\mu$m), characterized by the Projected Rayleigh Statistic $Z_{\mathrm x}$ vs $\log(N_{\mathrm H})$ for four sub-regions of the Vela\,C cloud \citep[][Fig. 5; MNRAS 474 1018]{soler2017,jow2018}. The dashed lines show linear fits to $Z_{\mathrm x} = C_{\mathrm Z}[\log_{10}(N_{\mathrm H}/cm^{-2}) - X_{\mathrm Z}]$; {(c)} Distribution of relative orientations of the gas filament with respect to $\hat{B}_{\textsc{pos}}$ as a function of $A_V$ in the Serpens South cloud \citep{pillai2020}. The NIR $\hat{B}_{\textsc{pos}}$ are shown as red filled circles and the FIR as blue filled circles. The magenta bar represents the lower $A_V$, first alignment transition suggested by Planck data, for the parallel-to-perpendicular transition near $A_V\sim$3.5 mag.}
    \label{fig:HRO-obs}
    \vspace{-0.5cm}
\end{figure*}

HROs indicate that the orientation between density structures in the ISM and the local magnetic field follows a bimodal distribution: below a certain column density, structures {preferentially} align {parallel to} the field, and above it, they are perpendicular to it (e.g., \citealt{Planck_XXXV2016}, see also Fig. \ref{fig:HRO-obs}). This observation points to a potential metric for the magnetic field strength with respect to gravity and turbulence, {discussed in detail in \S %\ref{sec:cloud_energy} and
\ref{sec:synth-HROs}}. }

\subsubsection{\textbf{Intensity gradients}}

The intensity gradient method \citep{koch2012, koch2012a, koch2013} estimates magnetic field strength from the measured angle between the magnetic field direction and the gradient in emission intensity, assumed to be representative of the resultant direction of motion of material due to magnetic, pressure and gravitational forces.  This method provides a point-to-point estimate of {both ratio of magnetic to gravitional and pressure energy and magnetic field strength,} and can be applied to any measure of plane-of-sky magnetic field direction.  {This method is applicable only where self-gravity is important \citep{koch2012}, but in these environments can probe possible evolution of the relative orientation of field/cloud structures \citep[e.g.][]{koch2014}, and the flow of material within filamentary structures \citep[e.g.][]{Anez2020,wang2020b}.}

\subsubsection{\textbf{Inclination angle}}
\label{sec:inclination}

{A challenge in studying the magnetic {field} properties of individual sources is that tracers are usually sensitive to either 
{$B_{\textsc{los}}$} (e.g.,~Zeeman splitting, Faraday rotation), or 
{POS} field morphology ({dust {polarization}}, velocity gradients), but not both.  For a large sample of objects {with} a random distribution of 3D magnetic field angles, the mean total magnetic field strength 
{$\langle B\rangle = 2 \langle B_{\textsc{los}}\rangle$}.
However, {for individual $B_{\textsc{los}}$ measurements, only a lower limit on $B$ can be set}.} 

{Magnetic field inclination angles also complicate the interpretation of dust polarization observations.  Clouds with weak magnetic fields tend to have less ordered field morphology (e.g. \citealt{ostriker2001, soler2013}, and \S \ref{sec:dcf}), and lower polarization levels due to signal cancellation as the polarization angle changes within the volume probed by a given sightline.  However, these observations are also consistent with viewing a more strongly magnetized cloud from an angle nearly parallel to the mean field direction \citep{king2018}, as any small variations in field direction would appear much larger when projected on the {POS}, while the polarization levels will be lower.  It is {thus} difficult to determine whether a cloud has a weak magnetic field or is viewed from a geometry where $\gamma$, the inclination angle of the magnetic field with respect to the {POS}, is large.}

Some studies have incorporated statistical methods such as Monte Carlo simulations and $\chi^2$ analysis to observations of both {POS} magnetic field morphology
and measurements of $B_{\textsc{los}}$ to model the 3D morphology of magnetic fields~\citep{Tahanietal2019}.
However this 
is complicated by the fact that different tracers are sensitive to the field in gas at different ranges of densities and gas phases, so each tracer may have individual biases {(see, e.g., \S \ref{sec:dcf_discussion}).}

{\cite{chen2019} developed a method to estimate the mean {magnetic field} inclination angle 
{from} polarized dust emission. Spinning 
{grains}
with their long axes perfectly perpendicular to the magnetic field should show no projected elongation if viewed parallel to the {field} ($\gamma\,=\,90^{\circ}$) and {so} no polarization \citep{hildebrand1988}. 
{The} projected {grain elongation} 
will be maximized when the {field} is 
{in} the {POS} ($\gamma = 0^{\circ}$).  If there are no variations in $\hat{B}_{\mathrm{POS}}$, or $\gamma$~along the {LOS} {then} 
measured fractional polarization $p$~is 
%\begin{linenomath*}
\begin{equation} \label{eqn:p_inc}
    p\,=\,\frac{p_0 cos^2\gamma}{1-p_0\left(cos^2\gamma-\frac{2}{3}\right)},
\end{equation}
%\end{linenomath*}
where $p_0$~is the intrinsic fractional polarization (assumed to be constant within the cloud). $p_0$~can be estimated from the maximum $p$~observed in the cloud
%\begin{linenomath*}
\begin{equation} \label{eqn:p_max}
    p_{\mathrm{max}}\,=\,\frac{p_0}{1-\frac{1}{3}p_0}.
\end{equation}
%\end{linenomath*}
Using Monte Carlo and synthetic observations of MHD simulations \cite{chen2019} show that estimates of the mean density weighted inclination angle $\hat{\gamma}^{2D}$ from eq.~\ref{eqn:p_inc} are biased towards intermediate {$\gamma$}.
They derived numerical correction factors from \textsc{Athena} MHD colliding flow simulations that can estimate of $\hat{\gamma}^{2D}$ to within 10--30$^{\circ}$ accuracy.
\cite{Sullivan2021} applied this method to nine polarization maps of molecular clouds from {\em Planck} and BLASTPol and found  
$\langle\gamma\rangle$ ranging from 16$^{\circ}$ (Musca/Chamaeleon), to 69$^{\circ}$ (Perseus).   Additional numerical and observational studies are needed to determine whether similar methods of estimating $\gamma$ can be applied to low resolution data, to both super- and sub-Alfv\'enic clouds, and to clouds with variations in dust temperature and grain alignment efficiency.}

{{
\subsubsection{\textbf{Ion-to-neutral linewidth ratio}}

\citet{Houdeetal2000a, Houdeetal2000b} showed that in molecular clouds, the linewidths of coexisting ions and neutrals {differ in the} presence of strong magnetic fields.
{Two effects are posited to cause this difference: firstly, that neutral linewidths trace turbulence, while narrower ion linewidths trace} gyromagnetic motion around {field} lines. {This difference may be used to probe 3D magnetic fields by determining the inclination angle $\gamma$ which,
combined with Zeeman and dust polarization observations, can describe the 3D field \citep{Houdeetal2002,houde2004}.}

{Alternatively,} \citet{LiHoude2008} {suggest} that the {narrower ion linewidth is due to the differing turbulent velocity dispersion spectra of neutrals and ions below the ambipolar diffusion size scale}.  {They propose
%\begin{linenomath*}
\begin{equation}
    B_{pos} = \left[\frac{L^{\prime}}{0.5\,{\rm mpc}}\frac{\sigma_{v,\textsc{nt},n}}{1\,{\rm km\,s}^{-1}}\left(\frac{n_{n}}{10^{6}\,{\rm cm^{-3}}}\right)^{2}\frac{x_{i}}{10^{-7}}\right]^{\frac{1}{2}}\,{\rm mG},
\end{equation}
%\end{linenomath*}
where 
$L^\prime$ is the ambipolar diffusion size scale, at which ions and neutrals decouple (cf. eq.~\ref{eq:ambipolar_diff}), and $\sigma_{v,\textsc{nt},n}$ is measured at $L^{\prime}$.
This method has been used to measure field strengths in dense regions \citep{Hezarehetal2010,tang2018}. However, recent observations of the dense core B5 have found an ion linewidth greater than that of the neutrals \citep{pineda2021}, suggesting more complex field dynamics at high densities.
Degeneracies between these two effects, and so between their measurements of $\gamma$ and $B_{pos}$ respectively, may exist \citep{Houde2011}.}}}

\section{\textbf{MAGNETIC FIELDS WITHIN MOLECULAR CLOUDS}}
\label{sec:clouds}

\FloatBarrier

{In this section we discuss the properties of magnetic fields within molecular clouds, with a particular focus on large scales ($\gtrsim$\,1\,pc), and lower density regions of clouds (n$_{\mathrm{H}_2}\lesssim$\,1000 cm$^{-3}$). While we do include a discussion of the role of magnetic fields in the formation of cloud substructure (\S \ref{sub:substructure}), and the influence of magnetic fields on the star formation efficiency within clouds (\S \ref{sec:sf_properties}), we leave a detailed discussion of the magnetic field properties of dense gas substructures to the following sections.}

{One bias that should be noted is that while most stars form in giant molecular clouds (GMCs) ($M_{
\mathrm{cloud}}\,>\,10^5\,M_{\odot}$) we do not have many observations of magnetic fields in the outer envelopes of GMCs.  Most GMCs, {with the exception of Orion A and B}, are within a few degrees of the Galactic plane, where line-of-sight confusion makes unambiguously mapping the magnetic {fields} of individual 
cloud envelopes challenging. For such clouds polarized dust emission can generally only probe magnetic fields in the high column density filamentary regions
(see, e.g., the discussion of IRDCs in \S \ref{sec:irdcs}). Similarly, determining the Faraday rotation measure contribution caused by an individual molecular cloud along a crowded line of sight is challenging. Only velocity resolved tracers, such as Zeeman splitting, can probe the magnetic fields of individual clouds along a crowded sightline. 
}

{Most of our understanding of magnetic fields on cloud scales therefore comes from maps of mostly low-mass nearby clouds that appear to be off the Galactic plane (e.g.,~the clouds studied by \citealt{Planck_XXXV2016}), or are fortuitously located along relatively uncontaminated sightlines, such as the Vela\,C molecular cloud \citep{fissel2016}.  Future studies using near-IR extinction polarimetry, e.g.,~from the Galactic Plane Infrared Polarization Survey (GPIPS) of stars at different distances \citep{Clemens2020}, or resolved observations of GMCs in nearby galaxies that are not observed edge-on are will be needed to better understand cloud magnetic fields and their relation to galaxy-scale fields.}

\subsection{\textbf{The structure of magnetic fields in and around molecular clouds}}
\label{sec:3DField}

{Cloud-scale ($>$1pc) dust polarization, Faraday rotation, and Zeeman splitting observations generally indicate that molecular clouds have an ordered magnetic field structure with a high degree of correlation on $\sim$10\,pc scales \citep{planck2016-XXXV, fissel2016,Tahanietal2018}.  These observations are generally found to be consistent with simulations where clouds are strongly magnetized ($\Malf\,\approx$\,1), while weaker-field simulations show disordered and tangled field structure \citep{li2015a}.}
{Dust polarization-derived $\hat{B}_{\textsc{pos}}$ maps of nearby molecular clouds show many examples of large-scale bends in the magnetic field direction projected onto the plane-of-the-sky \citep{planck2016-XXXV} (e.g., Fig. \ref{fig:orion_a}). This may indicate that the magnetic field direction has been altered by interactions between the clouds and their environment.}

{
The $B_{\textsc{los}}$ observations of the Orion A, California, and Perseus clouds find that magnetic fields tend to point toward us on one side of these filamentary GMCs and away from us on the other side~\citep{Tahanietal2018}, i.e., $B_{\textsc{los}}$ reverses direction across the cloud along the filament's short axis, as shown in Fig.~\ref{fig:orion_a}.
This coherent $B_{\textsc{los}}$ reversal in GMCs indicates a structured magnetic field morphology associated with these clouds. Using Monte-Carlo simulations and considering systematic biases between the $B_{\textsc{los}}$ and $B_{\textsc{pos}}$ observations, \cite{Tahanietal2019} studied the 3D morphology of magnetic fields associated with  Orion A  and found that an {arc}-shaped\footnote{{Sometimes referred to as bow-shaped}; pronounced /b\={o}/ as in rainbow} magnetic field is the most probable candidate to explain the observed $B_{\textsc{los}}$ reversals in this region ($\sim 50$\,pc scale), as shown in the inset of Fig.~\ref{fig:orion_a}.
We note that some $B_{\textsc{pos}}$ observations on smaller (sub-parsec) scales near dense filaments {suggest} a helical morphology~\citep[e.g.,][]{Poidevinetal2011,Alvarezetal2021}, while this {arc}-shaped morphology has been observationally associated with larger structures thus far. 
{Arc}-shaped morphology is consistent with predictions of {{some}} cloud-formation scenarios, and it has been observed in ideal MHD simulations by \cite{Inoueetal2018} and \cite{LiKlein2019}. Although both of these simulations study regions on smaller scales ($\sim 4$\,pc), their results are applicable to larger scales ($\sim 50$\,pc). 

Moreover, mapping the 3D density structure of the ISM~\cite[e.g.,][]{Grossschedletal2018, Zuckeretal2018, Zuckeretal2019, Zuckeretal2020} can enable us to better determine the 3D magnetic field morphologies of molecular clouds. For example,  \cite{Grossschedletal2018} found that the `tail' of the Orion A cloud is inclined along the line of sight with a 70$^{\circ}$ inclination angle.  {The} presence of sheets or bubbles in the foreground and background of this cloud, as suggested by \cite{Rezaeietal2020}, strengthens the conclusion of an {arc}-shaped magnetic morphology for Orion A.  
We note that this {arc}-shaped structure in \cite{Tahanietal2019} is an approximate smoothed magnetic field morphology for the entire molecular cloud, and since the study focuses on estimating the overall coherent morphology, smaller field variations or observational effects are not resolved.  

Furthermore, 
{\citet[][and subm.]{Tahanietal2022Pers}} used $B_{\textsc{los}}$ observations and Galactic magnetic field (GMF) models, along with 3D cloud morphologies and $B_{\textsc{pos}}$ data, to reconstruct the complete 3D morphology and direction of their {arc}-shaped fields. This enabled them to find the large-scale plane-of-sky magnetic field directions in the Perseus and Orion A clouds, including the signed direction (without 180$^{\circ}$ ambiguity). They also found that the Perseus and Orion A clouds retain ``memory'' of the GMF, while some studies suggest that the Galactic and molecular cloud magnetic fields are decoupled from one another~\citep[e.g.,][]{Stephensetal2011}. 
In the Perseus cloud, they found that the coherent component of GMF, modeled by \cite{JanssonandFarrar2012} has the same orientation as the $B_{\textsc{pos}}$ observations. In 
Orion A,
they suggested that if
{only plane-of-sky measurements are considered,}
then the GMF appears parallel to the cloud, while the $B_{\textsc{pos}}$
{seen} by {\em Planck} {is}
perpendicular to the cloud. However, Orion A appears to retain a memory of the GMF if the 3D morphologies of the cloud, the GMF, and the cloud’s magnetic field are all considered.

A likely explanation for formation of an {arc}-shaped magnetic field morphology is the interaction of the field lines with the cloud's environment~\citep{heiles1989}. 
Feedback effects, such as supernovae explosions or expansion of \HII\ regions, can influence the magnetic field morphologies~\citep{heiles1989, soler2018, Tahanietal2019}. 
Moreover, observational and theoretical studies suggest that \HII\ regions can influence the magnetic fields and alter the field morphology of their parental clouds locally {on} smaller scales, resulting in magnetic field lines tangential to \HII\ region boundaries~(\citealt{Krumholzetal2007b, Santosetal2014, fissel2016, pattle2018, Dewangan2018, Konyves2021, Devaraj2021}; see also Fig~\ref{fig:NGC6334}e). 
}

\begin{center}
    \includegraphics[width=0.4\textwidth, trim={0.1cm .25cm 0cm .26cm},clip]{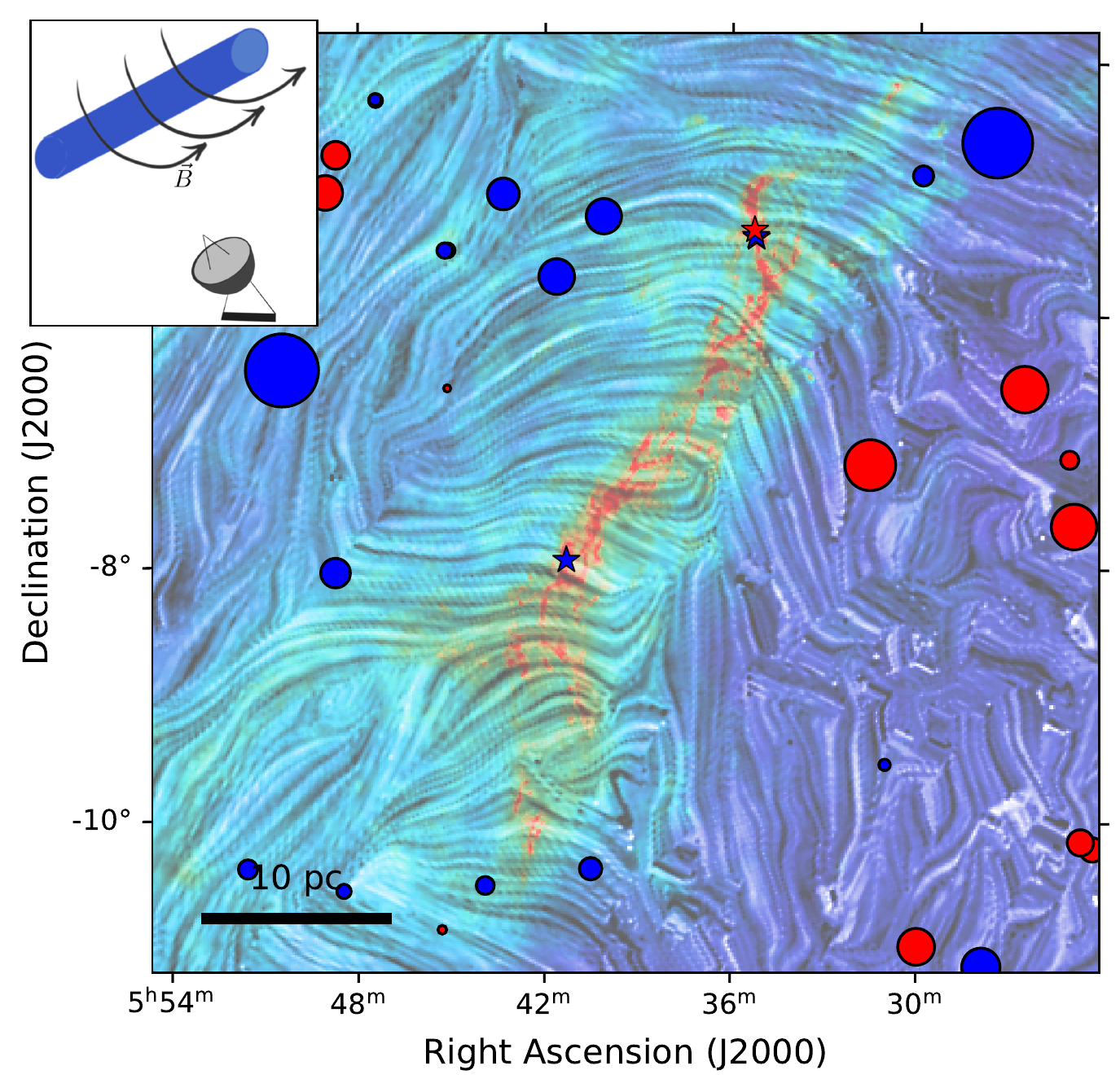}
    \captionof{figure}{Line-of-sight magnetic field of Orion\,A. The circles and stars depict $B_{\textsc{los}}$ {obtained by \cite{Tahanietal2018} and \citet[][OH Zeeman]{crutcher2010}, respectively, with blue (red) pointing toward (away from) us.} The drapery pattern illustrates $Planck$-measured $B_{\textsc{pos}}$ overlaid on a $Herschel$/$Planck$ column density map \citep{lombardi2014}. The size of the circles indicate the magnitude of $B_{\textsc{los}}$.
    The most probable explanation for these observations is an {arc}-shaped magnetic morphology \citep{Tahanietal2019}. }
    \label{fig:orion_a}
\end{center}

\subsection{\textbf{Formation of molecular clouds}}
\label{sub:formation}

{The formation of molecular clouds requires the accumulation and condensation of large quantities of diffuse ISM in a small volume, which various mechanisms can achieve. Here we examine each of these processes regarding their predictions for magnetic field morphology. Fig. \ref{fig:cloud_formation} provides a visual reference for each mechanism.

\subsubsection{Gravitational instability of the galactic disk}
\label{subsub:grav}

The interplay between gravity and magnetic buoyancy results in the Parker instability \citep{Parker66, Mouschovias74, shu74}. In this picture, the magnetic field lines are initially parallel to the disk, stabilizing it against gravitational collapse. However, gas volumes can become buoyant due to thermal feedback or cosmic ray propagation (a crucial contributor to the instability -- see \citealt{rodrigues16,Heintz2020}) and rise into the halo, carrying magnetic field lines along. This motion creates bends in the magnetic field and allows the formation of dense structures predominantly in `valleys' of converging magnetic field lines (see Fig. \ref{fig:cloud_formation}). 
The most unstable mode of the instability has a wavelength of $\sim$1~kpc, with a growth rate of the order of the Alfv\'{e}n crossing time \citep{Parker66,rodrigues16,HZ2018}. For a disk scale height of 100pc, and an Alfv\'{e}n speed $v_A=10{\rm km}/{\rm s}$, $\tau_A=h/v_A=10$ Myr.
(However, smaller-scale modes can grow faster in the non-linear regime -- see \citealt{Heintz2020}).
In differentially rotating disks, the Parker instability forms filamentary clouds with the magnetic field perpendicular to their main axis \citep{chou2000, Kortgen2018, Heintz2020}. These clouds, even if initially magnetically sub-critical, quickly become supercritical \citep{Kortgen19}.

\subsubsection{Condensations from large-scale turbulence}
\label{subsub:turb}

The turbulence driven by differential rotation and clustered stellar feedback creates shocks, triggering thermal instability and local collapse. This process involves a vast range of scales, posing a significant challenge for numerical models.
However, a novel technique of gradually zooming into clouds from a kpc-scaled box has recently allowed high-resolution studies of molecular clouds while preserving the large-scale dynamics of turbulence and magnetic fields. 
Examples include the models of \citet{walch2015} (SILCC), \citet{IbanezMejia2016},  \citet{kim2017} (TIGRESS), and \citet{hennebelle2018} (FRIGG).
Due to this setup's complexity, there is no single prediction regarding the shape of the magnetic field around the formed clouds. However, \citet{Girichidis2018} report that 
clouds in SILCC accrete preferentially along {field} lines.

\subsubsection{Colliding atomic flows}
\label{subsub:collflows}

When warm atomic flows converge to a shock, a series of fluid instabilities can condense the atomic gas into molecular clouds \citep{heitsch08,hennebelle08, VS11}: the Non-linear Thin Shell Instability (NTSI) enhances perturbations perpendicular to the shock surface, creating shear within the shock. The shear transitions to turbulence via the Kelvin-Helmholtz instability, and the condensations within the shock become thermally unstable, forming clumps of dense gas. The above scenario occurs in many situations, such as colliding shocks or the passage of spiral arms, so numerous numerical studies evoke it for molecular cloud formation. 

The magnetic field in this scenario can have a dominant role because it suppresses the relevant fluid instabilities. For instance, \citet{kortgen_2015} found a magnetic field $B>3\mu$G suppresses star formation. \citet{zamora-aviles_2018} showed that the magnetic field inhibits the growth of the NTSI, leading to more massive, denser, less turbulent clouds, with higher star-formation activity as the magnetization increases. \citet{sakre2020} confirmed this effect in colliding cloud simulations that track dense cores.

The orientation of the magnetic field is also a fundamental parameter in these experiments.  In general, an increasing inclination of the field with respect to the flows can delay the onset of dense gas formation \citep{InoueInutsuka2009, kortgen_2015}.
\citet{Iwasaki_2019} found that there is a critical angle $\theta_{cr}$ above which magnetic pressure completely suppresses the formation of molecular gas. The same study included an analytic estimate of $\theta_{cr}\le 15^{\circ}$ for a magnetic field $B>1\mu G$.
This conclusion has important implications for the allowed magnetic field morphology around molecular clouds. In colliding flow simulations, the magnetic field can only be primarily perpendicular to the flow collision interface. However, the magnetic field morphology within the slab and the filaments depends sensitively on the $\mathcal{M}_{A}$ of the flow collision.

\subsubsection{Shell expansion and interactions}
\label{subsub:shells}

The expansion and interaction of spherical shells, such as HII regions and superbubbles, is a particular case of colliding flows that has received much attention over the last decades \citep{Elmegreen_Lada1977,whitworth1994, McCrayKafatos1987, Tenorio-Tagle1987, Ehlerova1997, EN11} since star-forming clouds commonly surround feedback regions \citep{Deharveng2005, Deharveng2009, Dawson2013}.
However, forming molecular clouds out of a single shell expansion is challenging.  One reason is that the timescales for forming molecular clouds out of the diffuse atomic medium significantly exceed the evolution time of a shell. This effect is amplified by the presence of a magnetic field, as noted above. Besides, an arrangement of molecular clouds in a shell could reflect the pre-existing structure of the shell's surroundings \citep{Walch2015b}.
A shell interaction may still be insufficient for forming GMCs out of the diffuse atomic gas. \citet{Dawson2015} found that hydrodynamic simulations of supershell collisions could not explain the properties of an observed cloud between two supershells, hinting at a pre-existing dense structure. \citet{Ntormousi2017} found that magnetization may suppress the dense gas formation around the shells altogether.

Considering these difficulties, the ``multiple collisions'' model proposed by \citet{Inutsukaetal2015} becomes an attractive alternative. In this scenario, the passage of a single shell creates CNM, and subsequent shell interactions bring it to a molecular state. 

{{\subsubsection{Comparisons with observations}
Predictions for velocity and magnetic field structures from these models can enable us to compare them with the available and upcoming observations. This will provide tools to further modify and improve on these cloud-formation models. 

For example, the multiple-collision  model of \cite{Inutsukaetal2015} provides predictions of the 3D morphology of magnetic fields associated with formed filamentary structures and their velocity structure. In their model, the shock-cloud interaction can bend the magnetic field lines around the formed filamentary molecular clouds (on scales $\sim 1-100$\,pc). This magnetic field bending, regardless of how it is formed, allows for more mass accumulation on to the filamentary structures, resulting in dense filaments. We discussed observations of this {arc}-shaped magnetic morphology in \S \ref{sec:3DField}.
Moreover, velocity observations by \cite{Arzoumanianetal2018} and \cite{Bonneetal2020}, in a filament within the Taurus cloud and in the Musca filament, respectively, match the velocity description of \cite{Inutsukaetal2015}. 
\cite{Braccoetal2020} showed that their POS magnetic field observations, presence of shells, and evidence for compressed magnetic fields in the Corona Australis molecular cloud were consistent with predictions of \cite{Inutsukaetal2015} and bubble expansions and interactions. 
{\citet{Tahanietal2022Pers} explore} velocity information, coherent Galactic magnetic field models, and the orientation of line-of-sight magnetic field reversals (see \S\ref{sec:3DField}) associated with the Perseus cloud, and find them consistent with the predictions of shock-cloud interactions. 

We note that other filamentary cloud formation scenarios can potentially predict this {arc}-shaped magnetic field morphology. 
{For example, while the linear phase of the Parker instability predicts bending of the field lines on kpc scales, its non-linear evolution also involves smaller scales. The MRI is also a good candidate for bending magnetic field lines around dense structures. However, due to the complexity of the Galactic ISM, identifying a single origin of the observed features is close to impossible -- particularly because several or all of these processes may be at work simultaneously.} Therefore, more detailed predictions from each of these models regarding velocities and magnetic field morphologies are required to further distinguish between the models and to study whether one model is more favorable for certain regions within the Galaxy. 
}}
}

\begin{figure*}[!t]
    \centering
    \includegraphics[trim=0 0 0 0, width=0.8\textwidth]{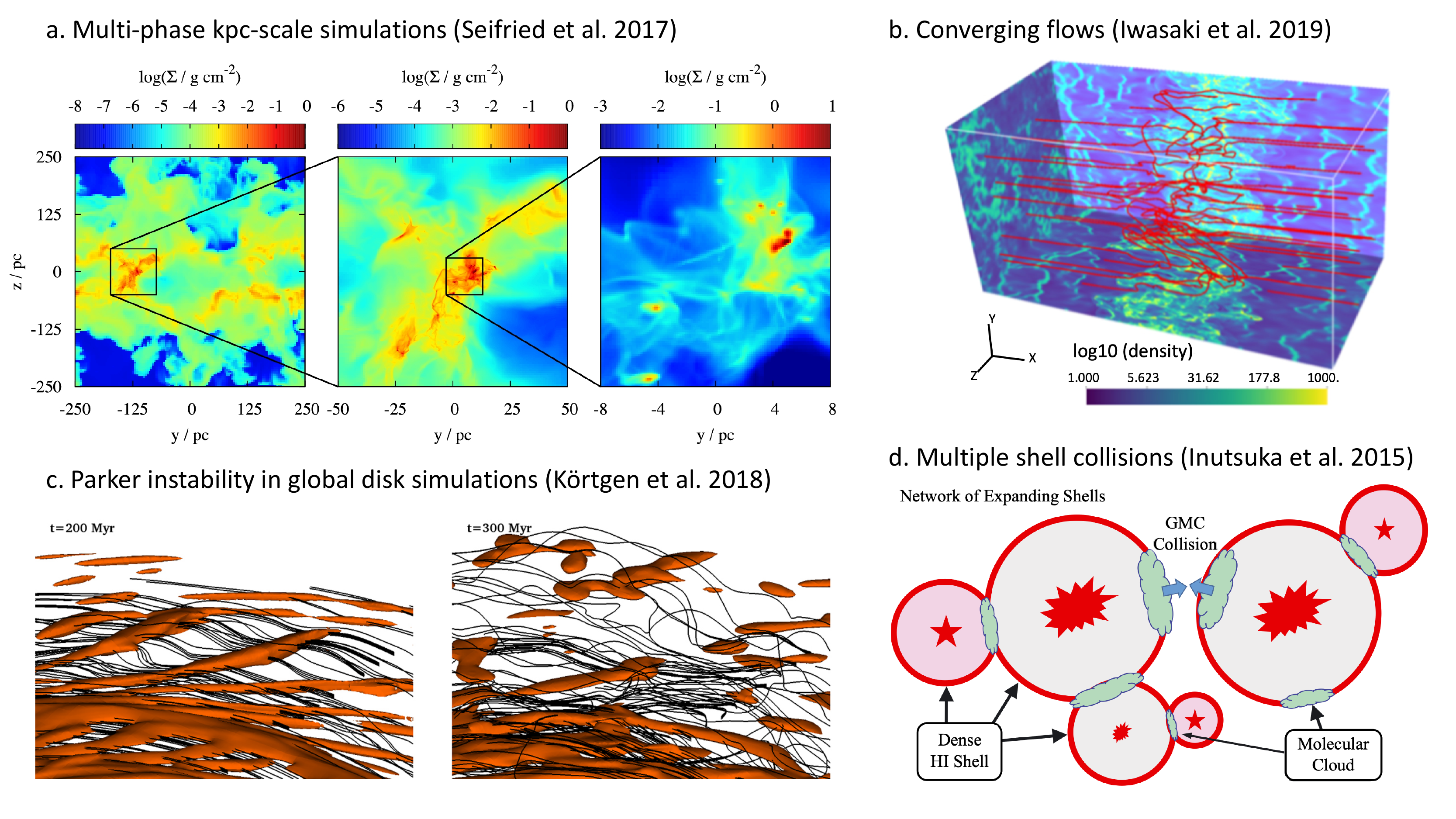}
    \caption{Cloud-formation mechanisms. (a) An example of the zoom-in technique in kpc-scaled simulations (\S \ref{subsub:turb}) from the SILCC simulations \citep[][Fig. 1; MNRAS 472 4797]{Seifried17}. Boxes {show} column density projections in the surroundings of a molecular cloud. (b) An illustration of {colliding flows}
    \citep[][\copyright\ AAS. Reproduced with permission]{Iwasaki_2019} (\S \ref{subsub:collflows}), showing density in slices and the magnetic field lines along the flows and at the collision interface. (c) A close-up on {diffuse clouds (brown surfaces show $10\,{\rm cm}^{-3}$ iso-density contours)} formed via the Parker instability in galaxy-scale simulations, showing magnetic field lines \citep[][Fig. 3; MNRAS 479 L40]{Kortgen2018} (\S \ref{subsub:grav}). (d) An illustration of the multiple shell collision model \citep[][reproduced with permission \copyright\ ESO]{Inutsukaetal2015} (\S \ref{subsub:shells}).}
    \label{fig:cloud_formation}
    \includegraphics[width=0.8\textwidth]{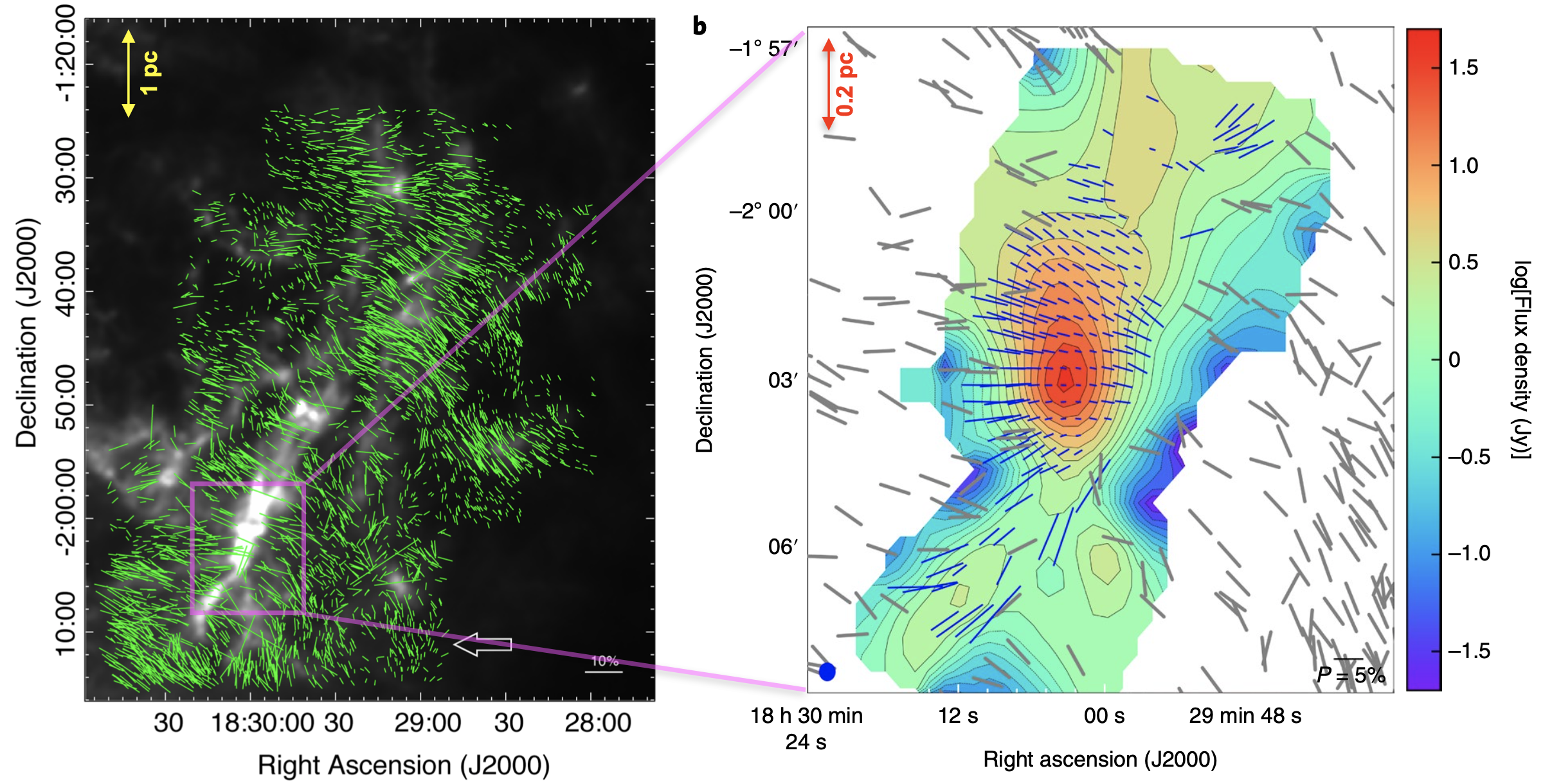}
    \caption{Left: $\textit{H}$-band polarization vector map of the Serpens South cloud superposed on {a} H$_2$ column density map \citep[][Fig. 6; PASJ 71 S5]{Kusune2019}. Right: NIR polarization vector map {(grey vectors; \citealt{sugitani2011})} and SOFIA/HAWC+ 214 $\mu$m polarization vector map (blue vectors), superposed on the SOFIA/HAWC+ 214 $\mu m$ intensity {map} \citep{pillai2020}.  }
    \label{fig:Serpens}
    \vspace{-0.5cm}
\end{figure*}

\subsection{\textbf{Energetic importance of magnetic fields}}\label{sec:cloud_energy}
\label{subs:energies}

{Molecular clouds are embedded in lower-density envelopes of mostly atomic hydrogen. Zeeman \HI~observations suggest this gas is both strongly sub-critical and has magnetic energy densities in approximate equipartition with turbulence \citep{Heiles2005}.   {\em Planck} observations of the alignment between observed magnetic field orientation and high-latitude filamentary structures are also broadly consistent with approximate equipartition between turbulence and magnetic fields in the diffuse ISM \citep{planck_2016_XXXII}.}

{\cite{Thompson2019}
{published} 38 OH absorption Zeeman measurements, which do not target dense sub-regions and therefore trace mostly lower-density molecular gas.  They find a mean {LOS} magnetic strength field $\langle B_{\textsc{los}}\rangle = 7.4\,\pm\,0.4\,\mu G$.  If the 3D field orientations of these sightlines are randomly distributed{, then} $\langle B\rangle\,\approx\,2\,\langle B_{\textsc{los}}\rangle$, implying the mean total field strength is {$\approx 15\,\mu$G}, which is larger than the $\langle B\rangle\,\approx\,6\,\mu$G derived by \cite{Heiles2005} for \HI.  No column density estimates, velocity line widths, or density estimates have been published yet for this sample so it is not possible to estimate the mass-to-flux ratio $\mu_{\Phi}$, or Alfv\'en Mach number $\mathcal{M}_{A}$.}

{Dust polarization observations are consistent with models where cloud-scale ($>$\,1\,pc) magnetic fields are dynamically important. \cite{Planck_XXXV2016} observations of 10 nearby clouds show a change in the preferred orientation of column density ($N_{\mathrm{H}}$) structures with respect to $\hat{B}_{\mathrm{pos}}$ from parallel to perpendicular as $N_{\mathrm{H}}$~increases (Fig. \ref{fig:HRO-obs}a), which is most consistent with simulations where the clouds are on average sub- or trans-Alfv\'enic. 
Similarly, DCF $B_{\textsc{pos}}$ estimates shown in Fig. \ref{fig:compilation} and discussed in \S \ref{sec:dcf_compilation} are mostly consistent with trans- or sub-Alfv\'enic gas motions (though as discussed in 
\S \ref{sec:dcf_discussion} these estimates tend to be systematically higher than Zeeman $B_{\textsc{los}}$ estimates at the same $n_{\textsc{H}}$). 
\cite{king2018} showed using \textsc{Athena} colliding flow simulations that the individual and joint PDFs of polarization fraction $p$ and local angle dispersion $\mathcal{S}$  are sensitive to both $\mathcal{M}_{A}$ and the mean field inclination angle $\gamma$.  Applying these techniques to {\em Planck} and BLASTPol observations of 9 nearby clouds, \cite{Sullivan2021} found the $\log(p)$ vs $\log(\mathcal{S})$ slope and $\langle \mathcal{S} \rangle$ better match {a} slightly super-Alfv\'enic {simulation}, {than a} sub-Alfv\'enic cloud simulation. {However, the authors note that if the calculation of  
{$v_{A}$} only considers the velocity component perpendicular to the field direction (rather than the 3D velocity), the super-Alfv\'enic simulation would be considered trans-Alfv\'enic.}}

{Line-of-sight velocity structure can also indicate the relative importance of gas turbulence and magnetic fields.  Thin spectral cube channel maps, where the velocity width is much smaller than the $\sim$several km/s turbulent linewidth of a GMC, are expected to be more affected by the turbulent velocity structure of the gas than the density structure \citep{Lazarian2000}. %Also cite Brunt and MacLow 2004?
\cite{Heyer2012} used principal component analysis of thin $^{12}$CO and $^{13}$CO spectroscopic channels in the Taurus molecular cloud to measure velocity anisotropy. They found a velocity anisotropy aligned with the magnetic field towards low column density cloud sightlines, from which they inferred that the outer cloud envelope of Taurus is sub-Alfv\'enic, while the dense gas structures are super-Alfv\'enic. \cite{Lazarian2018} used synthetic observations of turbulent MHD simulations to infer $\Malf$ from the PDFs of thin velocity channel gradient orientations.  Applying this method to $^{13}$CO observations of five molecular clouds{,} \cite{Hu2019} estimated mean cloud $\Malf$ ranging from 0.6 to 1.3.}

{Note that the observations discussed in this section show evidence of significant scatter in the estimated values for the mean Alfv\'en Mach number $\langle\Malf\rangle$ of individual clouds and within different sub-regions of molecular clouds \citep[e.g.,~][]{Hu2019, heyer2020}.  They are therefore consistent with molecular clouds having near equipartition between turbulence and magnetic fields, but with localized regions of sub- and super-Alfv\'enic gas motions.}

{At higher densities ($n_{\mathrm{H}}\,\gtrsim 1000 $\,cm$^{-2}$), OH and CN Zeeman measurements show that the maximum $|B_{\textsc{los}}|$ magnetic field strength begins to increase rapidly with $n_{\textsc{h}}$ (\citealt{crutcher2010, crutcher2012}; see also Fig. \ref{fig:compilation} upper left panel).   The exact density of the transition from a fairly flat distribution of $|B_{\textsc{los}}|$ vs. $n_{\textsc{h}}$, to a power-law increase has been the subject of some controversy (see discussion in \S \ref{sec:synth-Bvsn}).  However most authors agree that the transition is due to the onset of gravitational collapse, which under conditions of flux freezing will increase the magnetic field strength \citep{mestel1966,li2015a,chen2016,IbanezMejia2021}. }

{
{Interestingly,} the column density at which $|B_{\mathrm{LOS}}|$ begins to show a power-law increase, {$\log_{10}(N_{\mathrm{H}}/$cm$^{-2})\sim22$, \citep{crutcher2019}} is roughly the same as the column density above which structures tend to align perpendicular to the magnetic field rather than parallel $\log_{10}(N_{\mathrm{H}}/$cm$^{-2}){\sim21.7}$, though the transition column density shows considerable variation from cloud to cloud and between different sub-regions \citep{Planck_XXXV2016,soler2017, Soler2019}.}
{The number density at which $|B_{\textsc{los}}|$ increases (\citealt{crutcher2010}: $n{_H}\,\sim\,300\,{\rm cm}^{-3}$) is also similar to the characteristic density where gas alignment changes from parallel to perpendicular to $\hat{B}_{\mathrm{pos}}$ in Vela C ($n_{\mathrm{H_2}}\,\sim 10^3\,$cm$^{-3}$; \citealt{fissel2019}), and the density above which nearby filaments observed with {\em Planck} ($n_{\mathrm{H}}\,>\,1200\,{\rm cm}^{-3}$; \citealt{Alina2019}) preferentially align perpendicular to the background magnetic field.}
{\cite{chen2016} analysed \textsc{Athena} colliding flow simulations and found that both transitions roughly correspond to the density where kinetic energy due to gravitational contraction becomes larger than the magnetic energy.  This interpretation agrees with \cite{soler2017b} who
found that 
the change in relative alignment is caused by convergent gas flows, such as gravitational contraction of gas into dense filaments.}

\subsection{\textbf{Magnetic fields and substructure formation}}
\label{sub:substructure}
{If a magnetic field is well coupled to the gas, and dynamically important ($\Malf \lesssim 1$), it should also influence the formation of cloud substructure.  Strong magnetic fields result in anisotropic turbulence (which seeds structure formation), and set a preferential gas flow direction parallel to the field while resisting compression in the direction perpendicular to the field lines. {Magnetic fields can also help shape and reinforce long filamentary structures. For example, \cite{LiKlein2019} show that a moderately strong magnetic field ($\mathcal{M}_{A}\sim1$) is crucial for maintaining long and thin filamentary clouds for a long period of time, $\sim$0.5 Myr.}

As discussed in the previous section, observations of molecular clouds show a clear statistical correlation between the orientation of cloud structure and the magnetic field morphology.  {
{In the last few years}, more and more optical/NIR starlight polarization observations \cite[e.g.,][]{kusune2016,santos2016,Kusune2019,Sugitani2019} as well as sub-mm dust continuum polarization observations \cite[e.g.,][]{pillai2015,Lihuabai2015,planck_xxxiii,cox2016,liu2018,Alina2019,Tang2019,fissel2019,Soam2019,pillai2020,Doi2020,arzoumanian2020} 
have observed the magnetic fields surrounding filaments. These observations show that dense filaments preferentially align perpendicular to the direction of the local magnetic field, while lower-density filaments or striations tend to align parallel to the magnetic field.

For example, toward the nearby Musca cloud ($d \sim$ 200 pc), \cite{cox2016} find that both the low $N_{\mathrm{H}}$ filaments or striations and $B_{\textsc{pos}}$ are oriented close to perpendicular to the high-density main filament,
{similar} to observations of the Taurus B211/213 filament system \citep{chapman2011, palmeirim2013}. \cite{cox2016} propose a scenario in which local interstellar material has condensed into a filament
that is accreting background matter along field lines through the striations. \cite{Kusune2019} find that the filaments in the Serpens South cloud are roughly perpendicular to the global magnetic field (see the left panel of Fig. \ref{fig:Serpens}). They speculate that the filaments are formed by fragmentation of a sheet-like cloud that was created through the gravitational contraction of a magnetized, turbulent cloud.}

\cite{Planck_XXXV2016} statistically quantified this change in alignment using the histogram of relative orientations (HRO) method described in \S \ref{sec:hro} on 10 nearby clouds.  Fig. \ref{fig:HRO-obs}a shows that low-$N_{\mathrm{H}}$ structures tend to align parallel to the local magnetic field. The degree of alignment with the magnetic field then decreases with increasing $N_{\mathrm{H}}$, before changing to preferentially perpendicular in most clouds above $log_{10}(N_{\mathrm{H}}/{\rm cm}^{-2})\approx21.7$.  Such transitions in relative orientation only occur in simulations where cloud fields are dynamically important (e.g.,~\citealt{soler2013, hull2017}; lower panel of Fig. \ref{fig:HRO-obs}a), as discussed in \S \ref{sec:cloud_energy} and \S \ref{sec:synth-HROs}.

{
 \citet{Alina2019} further analysed the relative orientations between filaments, embedded clumps, and background magnetic fields for a sample of 90 {\em Planck} Galactic Cold Clumps (PGCCs) embedded in filaments where the background magnetic field orientation is uniform. They find that relative orientations between the filaments and their background magnetic field depend 
 on the contrast in $N_{\mathrm{H}}$ between the filaments and their background environment. In low-density ($N_{\mathrm{H,bkg}}<1.2\times10^{21}$ cm$^{-2}$) environments, low-density contrast ($\Delta N_{\mathrm{H}}<4\times10^{20}$ cm$^{-2}$) filaments preferentially have a parallel relative alignment with the background magnetic field, however, high-contrast ($\Delta N_{\mathrm{H}}>4\times10^{20}$ cm$^{-2}$) filaments show no preferred orientation. Interestingly, {PGCC-identified} filaments embedded in dense background environments ($N_{\mathrm{H,bkg}}>1.2\times10^{21}$ cm$^{-2}$) do not show {any preferential  orientation relative} to the background magnetic field. 
 In addition, filaments with densities larger than $\sim$1200 cm$^{-3}$ are mostly perpendicular to the background magnetic field \citep{Alina2019}.}

{Using polarization data at 250, 350, and 500 $\mu$m obtained by
{BLASTPol}, \cite{soler2017} found that the relative orientation between gas column density structures and the magnetic field changes progressively with increasing gas column density in the filamentary Vela C giant molecular cloud (see Fig. \ref{fig:HRO-obs}c). They find that the transition of the relative orientations depends strongly on the shape of the column density probability distribution functions (PDFs). The two regions with prominent power law tails in the column density PDFs have the clearest transitions from parallel to perpendicular alignment.  This could indicate that in regions where the change in orientation is prominent, the initial flows that created these regions were aligned close to the magnetic field direction, allowing dense gas to form efficiently without significantly increasing the magnetic flux.}

{\cite{Soler2019} later analyzed the relative orientations of structures in 36$^{\prime \prime}$~FWHM {\textit{Herschel}} column density maps, relative to 10$^{\prime}$ FWHM resolution {\em Planck} 353\,GHz maps of inferred ${\hat{B}_{\textsc{pos}}}$ for the 10 nearby (d $<$ 450 pc) clouds previously studied by \cite{Planck_XXXV2016}. In contrast to \cite{soler2017}, \cite{Soler2019} found that in cloud sub-regions with the steepest $N_{\mathrm H}$~power-law tail slopes {(power-law index $\alpha \gtrsim 2$)}, which are usually interpreted to indicate a region where the energetics are mostly turbulence rather than gravity dominated, the high-$N_{\mathrm H}$~structures tend to be aligned perpendicular to the magnetic field. In contrast, regions with the shallowest high-$N_{\mathrm H}$ power-law slope, which are generally thought to be the result of gravitational collapse of high density gas, have a mean alignment angle between $N_{\mathrm H}$ and $B_{\textsc{pos}}$, $\langle \phi \rangle$, closer to zero, indicating preferentially parallel alignment.  These results suggest that the relationship between the cloud/B-field may be more complicated than was inferred by \cite{soler2017}.  \cite{Soler2019} suggest that clouds with steep power-law slopes could represent sub-critical clouds, where strong magnetic support inhibits the formation of a high $N_{\mathrm H}$~power-law tail \citep{auddy2018}. However, this steep $N_{\mathrm H}$ power-law tail sample includes Orion A, the most active star forming region within 500\,pc distance, which is unlikely to be subcritical.

The transition in magnetic field vs.~cloud structure alignment also depends on gas volume density. \citet{fissel2019} compared the magnetic field orientation for the Vela C cloud inferred from 500 $\mu$m BLASTPol polarization maps to the orientation of elongated structures in Mopra integrated line intensity maps for nine different molecules. They find that 
the transition from parallel to no preferred/perpendicular alignment occurs between the densities traced by $^{13}$CO and by C$^{18}$O, which they estimate to be $n_{{\rm H}_2}\sim$10$^3$ cm$^{-3}$ \citep{fissel2019}.  This is similar to the transition density found by \citet{Alina2019} in {the nearby dense filaments found in the PGCCs catalog}, and to the $\hat{B}$ vs.~$n$ transition density for some large scale simulations (e.g.,~\citealt{seifried2020}). }

{Simulations also show that magnetic fields can influence the formation of dense filamentary structures.} {
\cite{Inoueetal2018} find that the shock compression of a turbulent inhomogeneous molecular cloud creates massive filaments, which lie perpendicular to the background magnetic field.  \cite{beattie2020} find that for cases with a strong magnetic field, corresponding to Alfv\'en Mach number $\mathcal{M}_{A}<1$, and turbulent Mach number $\mathcal{M}<4$, the anisotropy in the column density is dominated by thin striations aligned with the magnetic field, while for $\mathcal{M}>4$ the anisotropy is significantly changed by high-density filaments that form perpendicular to the magnetic field. The strength of the magnetic field appears to control the degree of anisotropy, but it is the turbulent motions controlled by $\mathcal{M}>4$ that determine which kind of anisotropy dominates the morphology of a cloud.}

Strong magnetic fields can also inhibit gravitational collapse by providing pressure support in the direction perpendicular to the magnetic field, which inhibits fragmentation.  This is further discussed in the section on filament fragmentation (\S \ref{sec:fil_fragmentation}).
In RAMSES simulations presented by \cite{Hennebelle2013}, hydrodynamical simulations without a magnetic field quickly fragment, while the filamentary structures that form in the MHD simulations remain more coherent, with the filaments confined by the Lorentz force.  A subsequent study by \citet{Ntormousi2016} that includes non-ideal MHD turbulence including ambipolar diffusion of neutrals with respect to the ions shows that such effects make the filamentary structures broader and more massive.  Note that these simulations did not include gravity. }

\subsection{\textbf{Correlations between cloud magnetism and star formation}}
\label{sec:sf_properties}

According to the theory of turbulent fragmentation, the distribution of stellar masses at birth (Initial Mass Function or IMF) 
{{is intimately connected to the Core Mass Function (CMF),which}}
mirrors the overdensity distribution of supersonic turbulence \citep{padoan97,hc08,hc09}. This hypothesis has led to the suggestion that isothermal, MHD turbulence might be sufficient to explain the observed {{peak}}
of the IMF \citep{haugbolle2018}, and in particular, its characteristic mass of $\sim$0.3 M$_\odot$. However, ideal, isothermal MHD turbulence imposes no characteristic scale, allowing filaments and cores to fragment up to the resolution limit (e.g., \citealt{Federrath2017, Lee2019}). On the other hand, several numerical experiments have reported little or no dependence of the shape of the IMF on magnetization \citep{Ntormousi2019, Guszejnov2020}, even in non-ideal MHD \citep{wurster2019}. Instead, \citet{Lee2019} showed that the dominant factor determining the shape of the IMF is the adiabatic high-density end of the equation-of-state. The magnetic field affects the peak IMF mass only when assigned unrealistically high values.

{Conversely,} numerical simulations show that magnetization plays a crucial role in setting the star formation efficiency {(SFE)} of molecular clouds (see \citealt{HIreview19} and \citealt{KFreview19} for extensive reviews).  Models of kpc-sized regions report suppression of the dense gas fraction and the overall star formation rate {(SFR)} of the model with increasing magnetic field strength \citep{Iffrig2017, Pardi2017, Girichidis2018}. Simulations of individual or colliding clouds \citep{wurster2019,wu2020}, massive, turbulent, star-forming clumps \citep{Myers2014}, and turbulent pc-sized GMC regions \citep{federrath2015} all show that with increasing magnetic field strength, the {SFR} decreases. 

However, there are only a few observational studies of this connection.
\cite{LiHB2017} investigated the correlation between magnetic field and star formation rate (SFR) in Gould Belt clouds. They argued that the clouds with a magnetic field predominantly perpendicular to their main axis consistently have lower SFR per solar mass than those with parallel alignment. However, \cite{Soler2019} found no evident correlation between the SFRs and the magnetic field orientation in the same clouds, leaving open the question of a possible correlation between magnetization and SFR.

\section{\textbf{MAGNETIC FIELDS INSIDE DENSE FILAMENTS}}
\label{sec:filaments}

\FloatBarrier

{

Thermal dust emission imaging surveys with the {\textit{Herschel}} Space Observatory have
discovered ubiquitous filamentary structures in nearby Giant Molecular Clouds
(GMCs) and distant Galactic Plane clouds \citep{Andre2010,andre2014,Schisano2020}. {\textit{Herschel}} observations also revealed that more than 70\% of prestellar cores and protostars are embedded in the densest filaments, with column
densities exceeding $\sim7\times10^{21}$ cm$^{-2}$, in nearby molecular
clouds \citep{andre2014,konyves2015}, strongly suggesting that dense filaments play a very important role in star formation. Numerical simulations have also found that magnetic fields are dynamically important in the formation of filaments as well as dense cores in molecular clouds (see \S \ref{sub:formation} and \ref{sub:substructure}). MHD simulations \citep{LiKlein2019} performed for the formation of large-scale filamentary clouds suggest a complicated evolutionary process involving the interaction and fragmentation of dense velocity-coherent fibers into chains of cores, resembling observations in nearby clouds, {such as in L1495/B213 \citep{Hacar2013}.} Observations of magnetic fields inside dense filaments \citep[M$_{line}\geq16~{\rm M}_{\odot}~{\rm pc}^{-1}$;][]{andre2014}, where the majority of dense cores and stars form, however, were very rare {a decade ago}.

\subsection{Magnetic field geometry inside {dense filaments and filamentary clouds}}

As discussed in \S \ref{sub:substructure}, observations indicate a trend that dense filaments preferentially align perpendicular to the direction of the local magnetic field. Magnetic fields inside dense filaments, however, are much more complicated than background magnetic fields due to interplay between magnetic fields, turbulence, gravity and stellar feedback \citep[see Fig. \ref{fig:NGC6334} for example;][]{arzoumanian2020}.

{In the last few years}, high-sensitivity and -resolution polarization observations with large single-dishes (e.g., JCMT, CSO, SOFIA) and interferometers (e.g., SMA, ALMA) have been resolving magnetic fields inside filaments at $<$0.1 pc \citep[e.g.,][]{Lihuabai2015,pattle2017a,Ching2018,Kock2018,cortes2019,pillai2020,Doi2020,arzoumanian2020,LiuJH2020,Guerra2021,Fernandez2021}. These observations are crucial for studying the roles of magnetic fields in the formation of dense cores and stars inside filaments.

\subsubsection{\textbf{Nearby filamentary clouds}} \label{sec:irdcs}

The {JCMT} B-fields In STar-forming Region Observations (BISTRO) survey has observed several filamentary clouds and revealed the magnetic field structures inside them. The first BISTRO polarization mapping of the OMC 1 region at 850 $\mu m$ found magnetic fields %lying 
{oriented} parallel to low-density, non-self-gravitating filaments, and perpendicular to higher-density, self-gravitating filaments \citep{wardthompson2017}. The densest region of the integral {shaped} filament in OMC 1 shows an hourglass field morphology, which is likely caused by the distortion of an initial field that is linear across the filament by the gravitational fragmentation of the filament and/or the gravitational interaction of clumps inside the filament \citep{pattle2017a}. \citet{Chuss2019} performed polarimetric observations of OMC 1 with SOFIA/HAWC+ at 53, 89, 154, and 214 $\mu m$. They find that at longer wavelengths (154 and 214 $\mu m$), the inferred magnetic field configuration
matches the `hourglass' configuration seen in previous
observations. However, the field morphology, differs at the shorter wavelengths (53 and 89 $\mu m$), specifically close to the Orion KL region because the short wavelength
data preferentially sample the warm dust that
corresponds to Orion BN/KL and the associated explosion, while the long-wavelength polarimetry is likely tracing the cooler outer part of the cloud \citep{Chuss2019,Guerra2021}.

In BISTRO survey data, the polarized emission from individual filamentary structures of NGC 1333 in the Perseus GMC is spatially resolved at 0.02 pc resolution
\citep{Doi2020}. The inferred magnetic field structure at 850 $\mu$m is 
{complex}, with each individual filament aligned at a different position angle relative to the local field orientation. Analysis combining the BISTRO data with low- and high- resolution
data derived from Planck and interferometers (CARMA) indicates that the magnetic field morphology drastically changes below a scale of $\sim$1 pc and remains continuous from the scales of {filament widths} ($\sim$0.1 pc) to that of protostellar envelopes ($\sim$0.005 pc or $\sim$1000 au). \cite{Doi2020} argued that the observed variation of the relative orientation between the filament axes and the magnetic field angles is mainly caused by {projection effects, and that
in 3D} space the B-field and the long axis of a filament are more likely perpendicular to each other.

NGC 6334, one of the nearest (d$\sim$1.3 kpc) filamentary clouds {forming} high-mass stars, has also been extensively studied in polarimetric observations at various scales (see Fig. \ref{fig:NGC6334}). From optical polarimetry and high angular resolution sub-mm polarization measurements on 
{100--0.01\,pc scales}, \citet{Lihuabai2015} found that there exist elongated gas structures
nearly perpendicular to the fields at all scales. However, the fields are symmetrically pinched near density peaks in many gas elongations (filaments or cores). 
{Using BISTRO data,} \citet{arzoumanian2020} revealed  
the characteristics of the small-scale  
($\sim$0.1 pc) magnetic field structure of the 10 pc-long hub-filament system {in NGC 6334}. 
They found variation in the field orientation and energy balance along the crests of sub-filaments. However, at smaller scales ($\sim$1 pc), the {POS} magnetic field ($B_{\textsc{pos}}$) angle varies coherently along the crests of the filament network. Along the sub-filaments that surround the densest ridge or hub structures, 
{$B_{\textsc{pos}}$} rotates from being mostly perpendicular or randomly oriented with respect to the crests to mostly parallel as the sub-filaments merge with the ridge and hubs. {\citet{arzoumanian2020} argue} that
this variation of the B-field structure along the sub-filaments may be caused by local velocity flows of infalling matter in the ridge and hubs.

\citet{pillai2020} also found a transition in relative orientation along the southern filament that {connects} to the hub region in the Serpens South cloud (see 
Fig. \ref{fig:Serpens}, {right panel}), that is, a return from perpendicular to parallel alignment at A$_V\sim$21 mag (see Fig. \ref{fig:HRO-obs}, {panel b}). They argue that this transition {may} be
caused by gas flow, indicating that gravitational collapse and star cluster formation can occur even in the presence of relatively strong magnetic fields.

Variation of the magnetic field structures surrounding or along filaments due to gravitational collapse or gas accretion is also seen {simulations}. 
{\citet{Zamora2017} performed 3D, self-gravitating MHD simulations, finding} that filaments (and subfilaments) {may form} by accretion/infall from the surrounding medium, driven by gravity rather than turbulence. 
{Material accretes} along the magnetic fields, which are oriented preferentially perpendicular to the filament skeleton. The magnetic field is at the same time dragged and bent by the velocity field due to the gravitational collapse. \citet{Gomez2018}, simulating molecular clouds undergoing global, multi-scale gravitational collapse, also find that the magnetic field is dragged by the collapsing gas in and around filaments. Around the filament, gas is accreted onto its skeleton and the magnetic lines are perpendicular to the skeleton.
However, as the gas density increases approaching the filament,
the gas flow changes direction, becoming almost parallel to the filament, and {field} lines are 
dragged to align with the filament. At the spine of the filament, however, {field} lines become perpendicular again since they must connect {to}
the opposite side of the filament, resulting in `U'-shaped magnetic structures, which tend to be stretched by the longitudinal flow along the filament. This picture is quite consistent with results from polarimetric observations of NGC 6334 \citep{arzoumanian2020} and Serpens South \citep{pillai2020}. 

\subsubsection{Distant filamentary clouds/clumps}

Magnetic fields inside distant massive filaments that may give birth to high-mass stars were barely known {a decade ago} due to observational difficulties: most of these clouds are located several kpc away on the Galactic Plane. 
{Since {then}, high resolution and high sensitivity polarization observations of distant and massive filamentary clouds have started to tackle this problem, 
revealing ordered magnetic field structures within them:  
the overall magnetic fields inside IRDCs G11.11-0.12 \citep{pillai2015}, G035.39-00.33 \citep{liu2018}, G34.43+0.24 \citep{Tang2019,Soam2019} and G14.225-0.506 \citep{Anez2020} are} perpendicular to the long axis of the main dense filaments.  However, such observations are still very rare.

\citet{pillai2015} studied magnetic fields inside two IRDCs, G11.11-0.12 and G0.253+0.016, using JCMT\slash SCUPOL. Magnetic fields inside G11.11-0.12 are perpendicular to the main dense filament, but are parallel to the lower density filament that merges onto the main
filament. In the G0.253+0.016 cloud, close to the Galactic center, the overall magnetic field morphology as well as the cloud morphology resemble an arched structure opening to the west, which is likely caused by strong shocks in this region. 

\citet{liu2018} observed the massive IRDC G035.39-00.33 with JCMT/POL-2 (see 
Fig. \ref{fig:Bfieldvel}, right panel). They found that the magnetic fields tend to be perpendicular to the densest part of the main filament. The magnetic fields, however, turn to become parallel to the main filament in the two ends of the main filament.  The magnetic fields also tend to be parallel to the low-density elongated structures that are connected to the main filament. The magnetic fields in the southern region of the main filament are likely pinched, hinting at an accretion flow along the filament or gravitational collapse of the massive dense cores therein.
The magnetic fields in the northern region of the main filament are parallel to the filament skeleton, which is likely caused by shocks induced by a cloud-cloud collision.  

Inside another massive filamentary cloud, G34.43+00.24, \citet{Soam2019} found that the core-scale ($\sim$0.1 pc) magnetic field lines seen with large single-dishes (JCMT\slash POL-2 and CSO) appear to be connected to the small-scale ($\sim$0.01 pc) field geometry traced by interferometers (CARMA, SMA) and large-scale ($\sim$10 pc) field lines traced by {\em Planck}. In the same cloud, \citet{Tang2019} found a close alignment between local magnetic field orientations and local velocity gradients (see Fig. \ref{fig:Bfieldvel}, left panel). This local correlation in alignment suggests that gas motions are influenced by the magnetic field morphology, or vice versa. 

\citet{wang2020a} observed a filament-hub system G33.92+0.11. In the high-density areas, their analysis shows that the filaments tend to align with the magnetic field and local gravity. In the low-density areas, they find that the local velocity gradients tend to be perpendicular to both the magnetic field and local gravity, although the filaments still tend to align with local gravity. 

However, 
{none of these single-dish observations resolved} magnetic fields down to 0.1 pc scale,
{preventing} comparative studies between these distant clouds and nearby clouds. 
{ALMA} will have adequate sensitivity/resolution to resolve magnetic fields down to $<0.1$ pc for these distant clouds in dust polarization or Zeeman observations with its mosaic mode.

\subsection{\textbf{The role of magnetic fields in filament stability and fragmentation}} \label{sec:fil_fragmentation}

{\cite{andre2014}} proposed a {paradigm} for star formation in which filaments play a fundamental role, based on {\textit{Herschel}} Gould Belt survey results. They argued that prestellar cores are formed through {gravitational} fragmentation of the densest filaments
above the thermally critical mass per unit length,  
$M_{line,crit}\approx16 M_{\odot}\,{\rm pc}^{-1}$ {(for a molecular gas temperature of $\sim$10 K)}. \cite{Andre2019} further demonstrated that the filament mass function (FMF) and the filament line mass function (FLMF) show very similar shapes that are both consistent with a Salpeter-like power-law function, {suggesting} that the stellar initial mass function (IMF) may originate from gravitational fragmentation of individual filaments.

The magnetic field, which is not much discussed in this paradigm \citep{andre2014,Andre2019}, could also play an important role in stabilizing filaments and regular filament fragmentation, as well as star formation inside filaments \citep{Nagasawa1987}. \cite{fiege2000} found that a poloidal field inside a filament helps to support the filament radially against self-gravity, increasing $M_{line,crit}$, while a toroidal field works with gravity in squeezing the filament, reducing $M_{line,crit}$. \citet{tomisaka2014} calculated magnetohydrostatic configurations of isothermal filaments that are laterally threaded by a magnetic
field, {finding} that the magnetic field 
{supports} the filament by increasing the maximum line-mass supported against self-gravity, and that the maximum mass of the magnetized filament is significantly affected by the magnetic field when the magnetic flux per unit length ($\Phi_{\rm cl}$) exceeds $\Phi_{\rm cl} \gtrsim 3\,{\rm pc\,\mu G}\,(c_s/190\,{\rm m\,s^{-1}})^2$.

\cite{Seifried2015} performed a set of 3D MHD simulations of magnetized filaments with various
{$M_{line}$ values and magnetic field orientations with respect to the major axis.}  They found that magnetic fields perpendicular to the major axis cannot contribute to the stabilization of supercritical filaments, resulting in filament widths $< 0.1$ pc. Conversely, a magnetic field parallel to the major axis can stabilize the filament against radial collapse, resulting in widths of 0.1 pc, in agreement with the observed filament width found in nearby clouds \citep{Andre2010,andre2014}. \cite{Seifried2015} also discovered three {filament} collapse modes:
edge-on, uniform, and centralized, depending on M$_{line}$. They found that filaments with M$_{line}$
equal to $M_{line,crit}$ (M$_{line,crit}\sim25~{\rm M}_{\odot}~{\rm pc}^{-1}$ for gas at 15 K) follow an edge-on collapse mode, with
star formation taking place at the outer edges of the filaments. 
No or only a little fragmentation is found along the major axes of these filaments. More and more fragmentation takes place along the entire filament (uniform collapse mode) for higher M$_{line}$. The filament collapses towards its common gravitational
centre (centralized collapse mode) if there is a initial moderate density enhancement in its centre (a factor of 3). 

Observations of some massive filaments indicate that magnetic fields are strong (ranging from several tens of $\mu G$ to several mG), and magnetic support is comparable to kinetic (thermal and turbulent) support in stabilizing filaments \citep{pillai2015,pattle2017a,liu2018,Tang2019,Soam2019,arzoumanian2020}. 

\cite{pattle2017a} derived a very high magnetic field strength ($B_{\textsc{pos}}=6.6\pm4.7$ mG) in the OMC 1 region. The magnetic energy density in OMC 1 is comparable to the gravitational potential energy density, suggesting that the
OMC 1 region is {on large scales} near magnetic criticality or slightly subcritical, with $\mu_{\phi,obs}\sim0.41$. {However, \citet{hwang2021} recently
found {the OMC 1 region to be} supercritical, with a median value of $\mu_{\phi,obs}\sim1.5$.
{This inconsistency is due to differing measurements of column density in the two works.}}
In NGC 6334, \cite{arzoumanian2020} find that in the outer parts of sub-filaments magnetic tension alone is not enough to balance gravity, while the inner parts of the sub-filaments seem to be in a magnetic critical balance. However, the combined magnetic and kinetic (thermal and turbulent) energies can provide sufficient support against gravity in these sub-filaments as well as in the densest ridge regions. 

\cite{pillai2015} find that the magnetic field in two distant IRDCs, G11.11-0.12 and G0.253+0.016, is strong enough to resist gravitational collapse and suppress fragmentation sufficiently to allow high-mass star formation.  The $\Malf$ values in the two clouds are low, $\mathcal{M}_{A}\leq1.2$, and the most likely values of $\mu_{\Phi,obs}$ are $<1$. Similarly, the three ridge regions (MM1, MM2, MM3) of the massive filamentary IRDC G34.43+00.24 also seem to be magnetically critical or slightly subcritical, with $\mu_{\Phi,obs}\sim0.5-1.1$ \citep{Tang2019}. \citet{liu2018} find that the main filament in IRDC G035.39-00.33, a cloud at a much younger evolutionary stage than G11.11-0.12 and G34.43+00.24, has a relatively weak magnetic field ($\sim50~\mu$G) and is likely unstable even if magnetic field support is taken into account.

Observational studies of fragmentation versus magnetic fields at all scales in filamentary clouds, however, are still very scarce. Toward IRDC G34.43+0.24, \citet{Tang2019} propose that the different fragmentation types seen at sub-parsec scale are determined by
{the varying relative importance of magnetic fields, gravity, and turbulence}. 
In regions where magnetic field dominates over turbulence, there is aligned fragmentation (MM2) or no fragmentation (MM1), while in MM3 where the magnetic field is not dominant over turbulence, clustered fragmentation is seen \citep{Tang2019}. \citet{Anez2020} 
{observed two hubs (Hub-N and Hub-S) in the IRDC G14.225-0.506 at 350 $\mu m$ using the Caltech Submillimeter Observatory (CSO)}.
They argued that different levels of fragmentation in these two hubs could be {caused} by magnetic field properties rather than gas density, because the density in the two hubs is similar.

{Over the last few years}, more and more high sensitivity and high angular resolution interferometric observations have also been performed to study the role of magnetic fields in the fragmentation process of distant massive filaments/clumps. With ALMA, \citet{Dallolia2019} observed the magnetic field in the filamentary high-mass star forming clump G9.62+0.19, and found that the magnetic field is oriented along the massive filament. The high magnetic field strength and smooth polarized emission in G9.62+0.19 indicate that the magnetic field could play an important role in the fragmentation of massive filaments and that the formation and evolution of dense cores can be magnetically regulated. ALMA dust polarization observations of the W43 high-mass star forming regions hint that a strong magnetic field may suppress fragmentation of clumps/filaments \citep{cortes2019}, and also control the angular momentum distribution from the core scale down to the inner part of the circumstellar disk where outflows are launched \citep{Arce2020}.  \citet{Palau2021} studied fragmentation and magnetic field within 18 massive dense clumps using the polarization data obtained in the {SMA} Legacy Survey of \citet{Zhang2014}. Their entire sample of massive dense clumps presents a strong correlation of the fragmentation level with the density of the parental clump. They also find a tentative trend of the fragmentation level with the mass-to-flux ratio.

{Observationally,} it is still hard to reach a firm conclusion on the role of magnetic fields in regulating filament fragmentation. However,
more high resolution dust polarization observations with state-of-the-art single-dishes (e.g., JCMT/POL-2, SOFIA/HAWC+) or interferometers (e.g., ALMA)
will allow deeper studies of magnetic fields inside {large samples of} filaments in the near future.}

\begin{center}
    \vspace{-0.5cm}
    \includegraphics[width=0.5\textwidth]{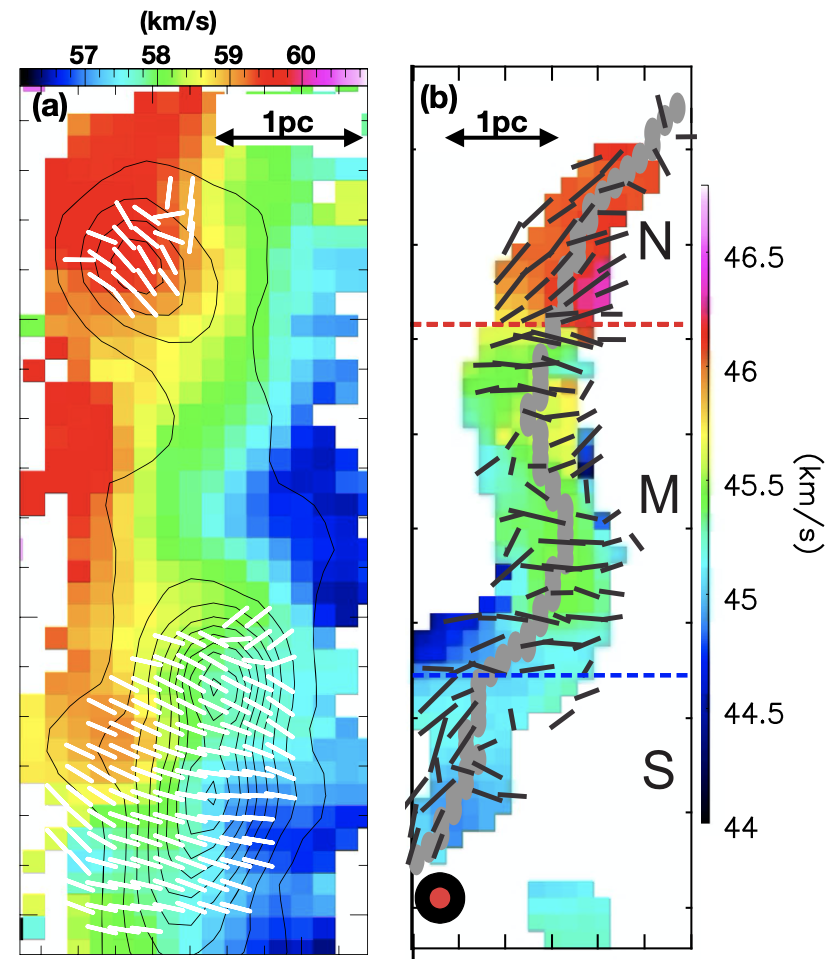}
    \vspace{-0.5cm}
    \captionof{figure}{Magnetic field vs. velocity field for two infrared dark clouds. Left: IRDC G34.43+00.24 \citep[][\copyright\ AAS. Reproduced with permission]{Tang2019}. Image shows gas velocity of N$_2$H$^+$ (1-0) line emission; contours show the integrated intensity map of N$_2$H$^+$ (1-0) line emission; white segments show magnetic field orientations inferred from polarized 350 $\mu m$ dust emission obtained with CSO. Right: IRDC G035.39-00.33 \citep[][\copyright\ AAS. Reproduced with permission]{liu2018}. The image shows gas velocity of NH$_3$ (1,1) line emission. The black segments show magnetic field orientations inferred from polarized 850 $\mu m$ dust continuum emission obtained with JCMT/POL-2.}
    \label{fig:Bfieldvel}
\end{center}

\section{\textbf{MAGNETIC FIELDS IN DENSE CORES}}
\label{sec:cores}

\begin{figure*}
    \centering
    \includegraphics[width=\textwidth,clip, trim={0cm 0.1cm 0cm 0cm}]{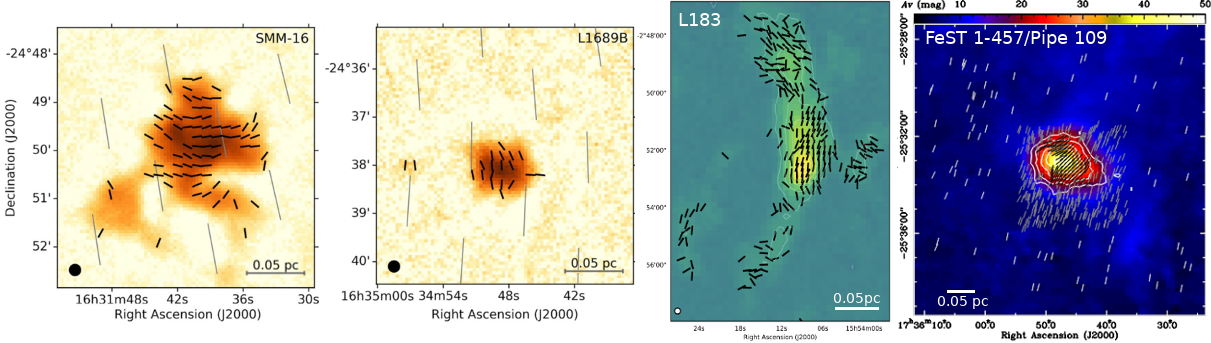}
    \vspace{-0.5cm}
    \caption{Recent observations of resolved magnetic fields in nearby low-mass prestellar cores.  {L-R: L1689 SMM-16 and L1689B \citep[][\copyright\ AAS. Reproduced with permission]{pattle2021}, L183 \citep[][\copyright\ AAS. Reproduced with permission]{karoly2020}, FeST 1-457/Pipe 109 \citep[][reproduced with permission \copyright\ ESO]{alves2014}.} All POL-2 observations (plus \textit{Planck} large-scale field for {L1689}) except FeST 1-457 (PolKa, large-scale field from extinction polarimetry).} 
    \label{fig:pscs}
    \vspace{-0.5cm}
\end{figure*}

{In this section we discuss magnetic fields within dense molecular cores.  We consider starless cores, which, if gravitationally bound (`prestellar'), will go on to form an individual stellar system \citep{benson1989,wardthompson1994} and low-mass protostellar cores, which have formed a central hydrostatic object, in \S~\ref{sec:lowmass_pscs}; their high-mass counterparts \citep{tan2014,motte2018} in \S \ref{sec:highmass_pscs}; 
and the coupling of protostellar outflows to their surroundings in \S\ref{sec:outflows_pscs}.  {We consider protostellar sources in their envelope-dominated main (Class 0) and late (Class I) accretion phases, rather than their later pre-main-sequence (Class II, III) phases \citep{lada1987,andre1993}.}  For a discussion of protostellar discs and outflows, 
see the chapter by Tsukamoto et al.}

{
Extinction polarization traces the peripheries of dense cores \citep[e.g.,][]{kandori2017a} while high surface-brightness cores are detectable in emission polarization \citep[e.g.,][]{wardthompson2000}, although the $A_{V}$ to which grains remain aligned
remains under debate \citep[e.g.,][]{alves2014, jones2015, pattle2019}, and depends on grain properties and incident radiation fields \citep{hoang2021}.  Starless cores typically lack internal structure and so are often resolved out by interferometers (e.g., \citealt{kirk2017}).  
Protostellar cores are typically easier to observe, being brighter and more centrally condensed, with an embedded source driving grain alignment, and with 
internal structure \citep[e.g.,][]{sadavoy2018}.  However, interpretation 
can be complicated by the
variety of {grain} alignment mechanisms around protostars (see \S~\ref{sec:dust}). 
Astrochemical analysis is therefore a key tool for investigating the dynamic importance of magnetic fields in dense cores.
}

\subsection{\textbf{Magnetic fields in low-mass cores}}
\label{sec:lowmass_pscs}

\subsubsection{\textbf{Magnetic field geometry}}

{The classic {dense} core magnetic field model is an `hourglass' morphology centred on the centre of mass, formed by material first collapsing freely along field lines, and the largely flux-frozen field then being dragged in by further collapse \citep{mestel1966};  \citet{ewertowski2013} and \citet{myers2018} present analytical 3D hourglass models.  The importance of ambipolar diffusion, and so the initial $\mu_{\Phi}$ value, may be inferred from the field morphology: stronger ambipolar diffusion leads to weaker flux-freezing, and so less field curvature \citep{basu2009}.

Submillimeter observations of starless cores show ordered, linear fields \citep{alves2014, liu2019, karoly2020, pattle2021}, without clear hourglass morphologies,
as shown in Fig.~\ref{fig:pscs}.  Magnetic fields are typically consistent with being parallel to the core minor axis in 3D \citep{basu2000}, suggesting that cores preferentially collapse along the magnetic field direction.  Simulations 
similarly produce triaxial cores, with the magnetic field preferentially most parallel to the core minor axis and most perpendicular to the major axis \citep{chen2018}. 

Parabolic models can be fitted to NIR observations of fields in starless core peripheries; SIRPOL observations of FeST 1-457 have been extensively modeled \citep{kandori2017a, kandori2017b,kandori2018a, kandori2018b,kandori2018c,kandori2020}, as have B335, B68,  BHR 71, CrA-E and L1174 \citep{kandori2020_B335,kandori2020_b68,kandori2020_bhr71,kandori2020_l1774,kandori2020_CrA-E}.  The parabolae fitted are in some cases significantly offset from the density peak; e.g., \citet{kandori2020_CrA-E}.
\citet{bino2021} fitted SIRPOL observations of FeST 1-457's periphery with an
{hourglass} model, finding that the core is mildly supercritical, and may have undergone a sub-to-supercritical transition through ambipolar diffusion.  However, as discussed in \S \ref{sec:cores_largescale} and shown in Fig.~\ref{fig:pscs}, magnetic fields in the centres of {starless} cores can be significantly offset from the preferred axis of the field in their surroundings, including in FeST 1-457 \citep[cf.][]{kandori2020}.} 

{Recently, \citet{Sahu2021} detected extremely high density ($n>10^7$ cm$^{-3}$) substructures inside five evolved prestellar cores in the Orion GMC with ALMA, {with} a spatial resolution of $\sim$320 au. Future high-resolution dust polarization observations toward these cores will shed light on how magnetic fields {behave} in the central regions of prestellar cores that are {developing substructure and} close to the onset of {protostar} formation. }

{At least some protostellar cores have hourglass fields when viewed using interferometers {\citep[e.g.][]{girart2006, maury2018}}, including protobinary cores \citep{kwon2019}.  {MHD collapse models can give a measure of initial $\mu_{\Phi}$ and field strength, and of collapse timescale, in these cores \citep[e.g.][]{frau2011}.}  However, ALMA has shown that many protostellar sources have more complex small-scale magnetic field geometries \citep[e.g.,][]{hull2017}, including potential magnetized accretion structures in 
{multiple} systems \citep[e.g.,][]{sadavoy2018}.  Interpretation of these observations may be complicated by polarized emission preferentially tracing outflow cavity walls \citep{hull2020}.  How common hourglasses are in protostellar cores remains uncertain.}

\subsubsection{\textbf{Magnetic field strength and energy balance}}

{{DCF} measurements of magnetic field strength in starless/prestellar cores range from $\sim 10^{1}$\,$\mu$G \citep[e.g.,][]{kirk2006} to a few $\times 10^{2}$\,$\mu$G \citep[e.g.,][]{karoly2020,pattle2021}, {(cf. Fig.~\ref{fig:compilation})}.  As discussed in \S\ref{sec:dcf_discussion}, isolated cores appear to have relatively low magnetic field strengths for their density; however, like other structures considered, their turbulence is typically found to be sub- or trans-Alfv\'enic.  DCF-derived mass-to-flux ratios in prestellar cores range from mildly supercritical (e.g., $\mu_{\Phi}\sim 2-3$; \citealt{kirk2006}) to somewhat subcritical (e.g., $\mu_{\Phi}\sim 0.1-0.4$; \citealt{karoly2020}).  A meta-analysis of DCF measurements of 17 nearby low-mass {pre- and protostellar} cores \citep{myers2021} suggests that these cores are nearly critical, with $B_{\textsc{pos}}\propto N^{1.05\pm 0.08}$ and $\propto n^{0.66\pm 0.05}$, and masses $\sim 0.5-2\times$ their virial mass.  They suggest that the cores' centrally-condensed and spheroidal geometries create the observed $\kappa\approx 2/3$, despite the cores' magnetic fields being as strong as possible while remaining subdominant.}  {However, \citet{Ching2022} present Zeeman measurements of the isolated prestellar core L1544, finding that both the core and its surrounding envelope are supercritical, suggesting that in at least some cases, prestellar cores may be supercritical from an early stage in their evolution.}

{{Astrochemical analysis} is a powerful tool for determining the magnetic criticality of cores.  The best tracers to use remain under investigation through chemical post-processing of non-ideal MHD {simulations} 
\citep[e.g.,][]{ferradachamorro2021}.  \citet{Priestley2019} suggest that HCN and CH$_3$OH line intensity profiles show promise as tracers of sub/super-critical collapse.  \citet{Yin2021} find the peak intensity ratio of N$_{2}$H$^{+}$ to CS lines, and the CS blue-to-red peak intensity ratio, to be useful diagnostics of the initial mass-to-flux ratio of cores.  The authors apply this analysis to the core L1498, identified as supercritical by \citet{kirk2006}, and suggest that it has evolved to supercriticality from a magnetically subcritical initial state.

Spatially-resolved mapping can provide information on core evolution; ambipolar-diffusion models predict an extended region of significant depletion of many species in the inner regions of cores \citep{tassis2012}, largely inconsistent with observations \citep{priestley2018}.  With assumptions, it may also be possible to use gas velocities to infer magnetic field strength in cores \citep{auddy2019}.}

\subsubsection{\textbf{{Prestellar core} lifetimes}}

{Core lifetime is a key measure of the dynamic importance of prestellar core magnetic fields.  A core in a state of unimpeded infall will collapse on its freefall timescale (eq. \ref{eq:t_ff}), while a core evolving quasistatically will have a lifetime set by the ambipolar diffusion timescale (eq. \ref{eq:t_ad}).  For typical core conditions, $t_{AD}$ is several times larger than $t_{ff}$, 
although the strong-field collapse time varies between models (e.g., $\gtrsim 5-10\times t_{ff}$; \citealt{machida2018}), and for sufficiently low values of $x_{e}$, $t_{AD}$ could become comparable to the freefall time, as discussed in \S\ref{sec:cores_ioniz}.}

{
Astrochemical studies have produced a range of results in different cores, often suggesting that chemistry in starless cores is not steady-state and that core lifetimes are $< t_{AD}$; e.g., \citet{pagani2013}, using N$_{2}$D$^{+}$/N$_{2}$H$^{+}$ as a chemical clock, ruled out slow contraction in L183, suggesting a core lifetime $\leq 0.7$\,Myr, while \citet{maret2013} modeled C$^{18}$O and HCO$^+$ in the cores L1498 and L1517B, finding ages of a few $\times 10^{5}$ yr.  Recently, \citet{bovino2021} studied the ortho-H$_2$D$^+$/para-D$_2$H$^+$ ratio in six cores in the Ophiuchus molecular cloud, finding core ages up to 200 kyr, consistent with $\sim 1-5\times t_{\rm ff}$, and inconsistent with $t_{AD}$.  However, at least some cores appear to be longer-lived: \citet{lin2020} found that the starless core L1512 is old enough for N$_{2}$ chemistry to be steady-state, hence $> 1.4$\,Myr, suggesting a lifetime $\gg t_{ff}$.  These estimates can be affected by pre-processing of core material: \citet{brunken2014} estimated the age of the IRAS 16293-2422 core as $>$1 Myr, while a reanalysis by \citet{harju2017} found an age of $\sim 0.5\pm 0.2$\, Myr, with $\sim 0.5$\,Myr pre-processing.}

{
Average core lifetimes can be estimated from number counts relative to Class II protostars (lifetime $\sim 2\,$Myr; \citealt{beichman1986}), assuming that all cores considered will go on to form stars, and that the local SFR
has remained constant over at least the Class II lifetime.  \citet{enoch2008}, from $Spitzer$ observations of nearby clouds, found an average lifetime of $\sim 500^{+500}_{-250}$ kyr
for cores with $n>2\times 10^{4}$\,cm$^{-3}$, indicating dynamic rather than quasistatic evolution, while \citet{kirk2005} found a mean lifetime $\sim 300$ kyr for SCUBA-identified prestellar cores.  For a large sample of cores, lifetimes can be estimated as a function of density, and are typically $\sim1-10\times t_{ff}$, decreasing with increasing density, suggesting that cores evolve from magnetically sub-critical at low densities to super-critical at high densities \citep{jessop2000,konyves2015}.  However, lower-density cores are less likely to be gravitationally bound, and so could be over-counted, inflating their lifetimes \citep{konyves2015}.  \citet{das2021} presented a partial flux-freezing model which produces lifetimes similar to the \citet{konyves2015} results, and suggest that cores are formed with a transcritical $\mu_{\Phi}$.}

{These results suggest that nearby, low-mass cores are generally evolving faster than the ambipolar diffusion timescale, but probably are not in 
{freefall collapse}.  Lifetimes are typically a few hundred kyr, with exceptions.  {The lifetimes of cores forming high-mass stars, and of low-mass cores in clustered environments, remain uncertain.}}

\subsubsection{\textbf{Ionization fraction in dense cores}}
\label{sec:cores_ioniz}

{For ion-neutral coupling to persist in dense cores and for magnetic fields to provide any support against collapse, the ionization fraction $x_{e}$ must be sufficiently high that $t_{AD} > t_{ff}$ \citep[e.g.,][]{caselli1998}; cloud-scale simulations show that $\mu_{\Phi}$ in dense cores increases with decreasing $x_{e}$ \citep{chen_2014}.  In starless cores, $x_{e}$ is maintained by cosmic ray ionization;
$t_{AD} < t_{ff}$ when $n_{H} \gtrsim 2.56\times10^{-12}x_{e}^{-2}\,{\rm cm}^{-3}$ \citep{caselli1998}.  \citet{caselli1998}, surveying 24 dense clouds, found $x_{e}\sim 10^{-6}-10^{-8}$, with cosmic ray ionization rate $\zeta\sim 10^{-16}-10^{-18}$\,s$^{-1}$, and that $t_{AD}\sim 50\times t_{ff}$ on average, although in 20\% their cores, $t_{AD}$ was only a few times greater than $t_{ff}$.  
The {highly centrally-condensed prestellar} core L1544 is particularly well-studied: 
\citet{caselli2002} found $x_{e} \sim 10^{-9}$, and so $t_{AD}\sim t_{ff}$.  \citet{keto2010} find $\zeta \sim 10^{-17}$ s$^{-1}$ in L1544, while \citet{ivlev2019} find L1544 is consistent with $\zeta\sim 10^{-16}$ s$^{-1}$, and \citet{bovino2020} measure 2--3$\times 10^{-17}$ s$^{-1}$.}

{Despite this, gas in molecular clouds may maintain a relatively high $\zeta$.  In the diffuse ISM, $\zeta\sim 10^{-16}$\,s$^{-1}$, decreasing with increasing $N$ \citep{indriolo2012}.  Fiducial values of cosmic ray ionization rate in dense gas are a few $\times 10^{-17}$\,s$^{-1}$ \citep{caselli1998,vandertak2000}.
However, values measured in molecular clouds can be considerably higher (e.g., $\sim 10^{-14}$\,s$^{-1}$ in OMC-2 FIR4; \citealt{favre2018}).  Shocks associated with protostars may produce cosmic rays within molecular clouds \citep{padovani2016}, and so $\zeta$ and $x_{e}$, and therefore $t_{AD}$ and the ion-neutral coupling in dense cores, may vary considerably both within and between clouds.}

\subsubsection{\textbf{Relationship between core- and cloud-scale fields}}
\label{sec:cores_largescale}

{\citet{chen2020} found the statistical orientation of cores 
with respect to the magnetic field of their parent cloud to vary: in the Perseus molecular cloud, the angles between core axes and background magnetic field are random, while in Taurus, cores are typically perpendicular to the background field, and in Ophiuchus, {under} large-scale feedback, cores are typically slightly parallel to the background field.

As shown in Fig.~\ref{fig:pscs}, significant discrepancies can exist between magnetic field directions in dense cores, and those in their immediate surroundings.  Comparisons of Planck and JCMT observations in nearby regions \citep{Doi2020, pattle2021, karoly2020}, and APEX and NIR observations of FeST 1-457 \citep{alves2014}, show that in some cores
the large- and small-scale fields agree, while in others
the two are perpendicular.  Bok globules also in some cases show discrepancies between their internal and external fields \citep{das2016, choudhury2019}.  Magnetic fields in cores embedded in filaments are often consistent with being perpendicular to the filament major axis, and so may be inherited from the parent filament \citep[e.g.,][]{pattle2021}, although there are exceptions \citep{eswaraiah2021}, and these observations beg the question of whether the field being traced is truly that of the core, or of the filament.
Simulations of ensembles of cores \citep{chen2018} suggest that internal and external magnetic field directions are typically correlated.  The number of cores with resolved magnetic field observations remains small, and it is not yet clear whether significant discrepancies between large- and small-scale core fields are unusual.}

\subsection{\textbf{Magnetic fields in high-mass star-forming cores}}
\label{sec:highmass_pscs}

{
Models of high-mass star formation fall into two categories: quasistatic core accretion, in which massive stars form analogously to low-mass stars, through monolithic collapse of a virialised high-mass prestellar core (HMPSC) \citet{mckeetan2003}; or competitive accretion, in which stars forming in a cluster acquire mass through either accretion of a common reservoir \citep{bonnell2006}, or accretion streams associated with hierarchical cloud collapse \citep[e.g.,][]{smith2009,vazquezsemadeni2017}.  At least some high-mass protostellar cores \citep{girart2009} and massive cluster-forming and fragmenting cores \citep{qiu2014,beltran2019} show hourglass magnetic field morphologies, suggestive of ordered collapse. 

Massive Dense Cores \citep[MDCs;][]{motte2007,motte2018} are a recent observational paradigm of high-mass star formation with no HMPSC phase \citep{tige2017}.  MDCs form within hub-filament systems, and have a starless phase, in which the MDC can contain numerous low-mass prestellar cores, followed by a star-forming phase, in which numerous low-mass protostars form, some of which become high-mass through competitive accretion from the MDC. 
MDCs appear to be collapsing on timescales $>t_{ff}$, with infall speeds $\sim 3-30$\% of their freefall velocities \citep{wyrowski2016}, indicating a potential role for magnetic fields in stabilising and mediating their collapse.

ALMA has observed a number of clumps and MDCs fragmenting into low-mass cores \citep[e.g.,][]{sanhueza2017, louvet2019}, and some fragmenting into cores with a wider range of masses \citep{raedelli2021}.  \citet{beuther2018} and \citet{sanhueza2019} identify samples of low-to-intermediate-mass cores in massive clumps.  \citet{beuther2018} find their cores to be near-virial and sub-Alfv\'enic based on DCF analysis, while \citet{sanhueza2019} find that 19\% of their 210 cores have masses $> M_{J}$.

{If there is an HMPSC star formation pathway, these HMPSCs will have masses $\gg M_{J}$, and so are expected to require significant magnetic support \citep[e.g.,][]{tan2014}. 
There is a short list of current HMPSC candidates (broadly defined as sources $\gtrsim 30\,$M$_{\odot}$, without internal structures and near virial equilibrium; e.g., \citealt{tan2014}),
G11.92-0.61-MM2 \citep{cyganowski2014}, a $\gtrsim 30\,$M$_{\odot}$ source with no signs of star formation; the $\sim\,60$\,M$_{\odot}$ core W43-MM1\#6 \citep[alternatively, a starless MDC;][]{nony2018}; and AG354-2 \citep{raedelli2021}.
Some intermediate-mass starless cores have also been identified, {e.g.} G331.372-00.116 ALMA1 \citep{contreras2018}, a $\sim$17\,M$_{\odot}$ infalling PSC candidate requiring a mG field for stability. }

Observations of resolved magnetic fields in high-mass star-forming clumps suggest magnetic supercriticality: \citet{pillai2016}, observing the massive star-forming clump G35.20w in W48, using SCUPOL and CN Zeeman data, found a highly ordered sub-Alfv\'enic magnetic field, but $\mu_{\Phi}\sim 1.5$.  \citet{LiuJH2020} observed three MDCs in the IRDC G28.34 with ALMA, finding the cores to be sub-virial, with magnetic fields well-aligned with local gravity, suggesting gravitational collapse.  However, a bimodal distribution of magnetic field/outflow axes suggests that magnetic fields could be dynamically important later in the collapse process.  Recent ALMA observations of the high-mass star-forming regions G327.3 \citep{beuther2020} and IRAS 18089-1732 \citep{sanhueza2021} have found toroidal magnetic field geometries, as shown in Fig.~\ref{fig:highmass_cores}, suggesting dynamics dominated by gravity and rotation, while
\citet{sanhueza2019} find $\mu_{\Phi}>8$ in IRAS 18089-1732, indicating a highly gravitationally-dominated system.

Models of turbulent massive cores suggest that magnetic fields affect fragmentation and multiplicity.  \citet{commercon2011} find that magnetic fields suppress fragmentation, suggesting that highly magnetised MDCs could produce isolated massive stars, while moderately magnetised MDCs form {small clusters}. Stronger fields may also reduce fragmentation in discs around massive protostars \citep[e.g.,][]{peters2011,mignonrisse2021}.  \citet{Palau2021}, using SMA data, find a tentative correlation between the number of embedded sources and $\mu_{\Phi}$ in a sample of MDCs, in agreement with models.

How the magnetic field geometry of a dense core changes under external feedback may be important to {core evolution} in massive star-forming regions.  There is limited evidence that linear fields {are} maintained in some cores under feedback \citep{cortes2021,pattle2021a,Konyves2021}, but more observations are {needed}.}

\begin{center}
    \vspace{-0.4cm}
    \includegraphics[width=0.5\textwidth,clip, trim={0cm 0.11cm 0cm 0.1cm}]{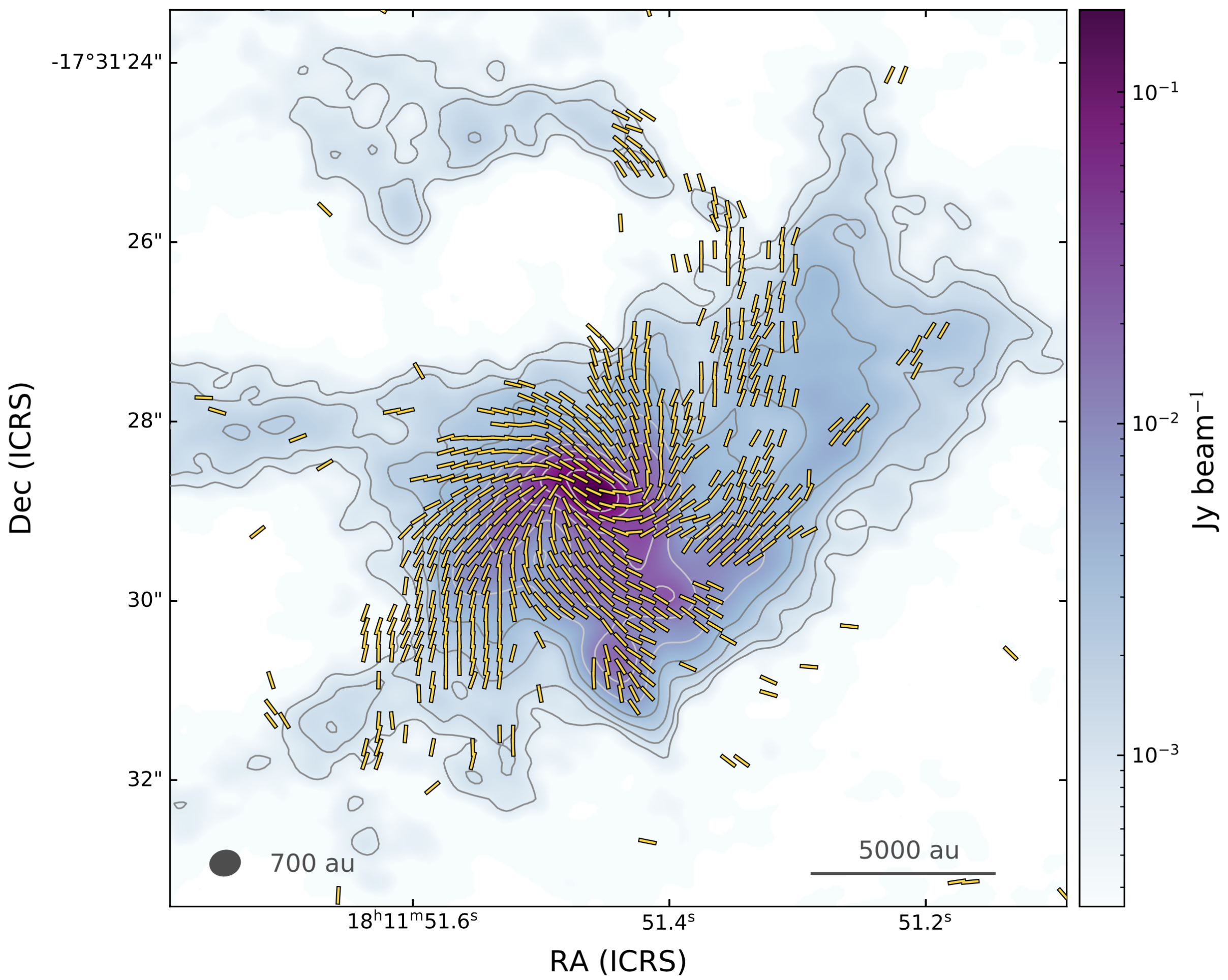}
    \vspace{-0.5cm}
    \captionof{figure}{ALMA 2.1mm observations of magnetic fields in the high-mass star-forming region IRAS 18089-1732 \citep[][\copyright\ AAS. Reproduced with permission]{sanhueza2021}.  The toroidal magnetic field geometry is suggestive of dynamics dominated by gravitational collapse and rotation.}
    \label{fig:highmass_cores}
\end{center}

\subsection{\textbf{Outflows and the effects of feedback}}
\label{sec:outflows_pscs}

{
Interferometric studies of core-scale magnetic field and outflow orientation have suggested that alignment is typically random, 
with a possible highly-correlated subset \citep{hull2014, Zhang2014,hull2019}, {while observations of protostellar envelopes find fields to be preferentially either parallel or perpendicular to the outflow, with misalignment more common in sources with larger rotational energies \citep{galametz2018,galametz2020}}.  A comparable single-dish study found that outflows and local magnetic fields tend to be misaligned by $50^{\circ}\pm15^{\circ}$ in 3D, but did not rule out random orientation \citep{yen2021}.
Alignment between protostellar outflows and larger-scale magnetic field or gas density structures may vary significantly between clouds. In G28.37+0.07, \citet{kong2019} found protostellar outflows to be preferentially perpendicular to the IRDC, suggesting a quite strong magnetic field.  Conversely, in Perseus, \citet{stephens2017} found outflow-filament angles consistent with either random alignment or a mix of projected parallel and perpendicular angles. }

{The initial magnetic field strength of a collapsing core may
{significantly affect its} protostellar outflows: in strong-field collapse, the angular momentum of the star-forming cloud may be efficiently transferred away before rapid collapse begins, resulting in weaker outflows \citep{machida2018}.
Magnetic fields couple such outflows to the surrounding cloud; these outflows 
provide {feedback on very short timescales ($\lesssim 10^{4}$ yr)}, switching on as soon as either a low- and high-mass protostar forms
\citep[e.g.,][]{KFreview19}.  We discuss this in \S \ref{sec:feedback}.}

\FloatBarrier

\section{\textbf{SYNTHESIS}}
\label{sec:synthesis}

We now return to the questions posed in \S \ref{sec:intro} and the methods described in \S \ref{sec:methods} to discuss our current understanding of how magnetic fields affect the star formation process across the range of size and density scales considered in \S\ref{sec:clouds}--\ref{sec:cores}.

\subsection{Energy balance}
\label{sec:eng_balance}

\subsubsection{\textbf{The reliability of DCF analysis}}
\label{sec:dcf_discussion}

We here discuss the {DCF measurement compilation} presented in \S~\ref{sec:dcf_compilation} {and} Fig.~\ref{fig:compilation}.  
Fig.~\ref{fig:compilation}b shows little systematic difference between DCF variants.
{We do not have space in this chapter to perform a fair and thorough comparison of the various DCF methods, and so we refer the reader to \citet{liu2021,liu2021a} for detailed discussion of this point.}

Fig.~\ref{fig:compilation}c shows continuity between extinction and emission measurements. 
Extinction measurements do not trace high $A_{V}$, while emission measurements 
{lack the pencil-beam resolution of extinction measurements,}
and so this continuity suggests {similar} magnetic field dispersion across a wide range of scales.  The high-density breakdown in the correlation between $B_{\textsc{pos}}$ and $n$ is exclusive to the interferometric points.
This could perhaps be due to instrumental effects (e.g., spatial filtering), or to DCF failing on small scales
\citep{liu2021}.  It is unlikely to be caused by $\sigma_{v,\textsc{nt}}$ tracing infall motions, as the $B_{\textsc{pos}}$ values decrease rather than increase, and this behavior appears to result from an increase in $\mathcal{M}_{A}$, which is independent of  $\sigma_{v,\textsc{nt}}$.
The apparent decreases in $B_{\textsc{pos}}$ are predominantly seen in distant ($>$ 1\,kpc) star-forming clumps.  Nearby protostellar sources and HH objects 
have $\mathcal{M}_{A}\sim 1$ to the highest densities.

Fig.~\ref{fig:compilation}d qualitatively suggests that isolated cores have lower $B_{\textsc{pos}}$ for a given $n_{\textsc{h}}$ than do structures under feedback.  However, there is no distinction in their $\mathcal{M}_{A}$ values (Fig.~\ref{fig:compilation}f); the differences appear to arise from the isolated sources having smaller $\sigma_{v,\textsc{nt}}$, and so lower $v_{A}$, (Fig.~\ref{fig:compilation}e).  `Cloud structures' is a broad category and 
covers a wide range of $B_{\textsc{pos}}$ values.
Protostars, jets, and HH objects have quite high $B_{\textsc{pos}}$ values, although these measurements are mostly at high densities where DCF is most uncertain.  Interestingly, the few extragalactic measurements appear to be broadly similar to those made within the Milky Way.

We measured the mean DCF- and Zeeman-derived magnetic field strength over each decade in density, and found that the $\langle B_{{\rm DCF}}\rangle/\langle B_{{\rm Zeeman}}\rangle$ ratio varies in the range 1--18, with an average of $B_{{\rm DCF}}/B_{{\rm Zeeman}} = 6.3\pm 1.5$, where the uncertainty is standard error on the mean (the median ratios produce a similar value).  {It is important to note that these compilations are not well-controlled or homogeneous samples, and that the DCF and Zeeman measurements may not have been made in the same environment at a given density. Moreover, measurements of gas density from paramagnetic species and from dust emission will themselves be subject to differing systematic uncertainties. However, this ratio} is similar to previous results: \citet{poidevin2013}, comparing CN and OH Zeeman to SCUPOL DCF measurements of matched regions, found $B_{{\rm DCF}}/B_{\rm Zeeman} = 4.7\pm2.8$, while \citet{soler2018}, comparing H\textsc{I}
Zeeman to $Planck$ DCF measurements of the
Eridanus superbubble, found $B_{\rm DCF}/B_{\rm Zeeman} \sim 2.5-13$.

However, {Zeeman} measurements trace $B_{\textsc{los}}$ while DCF traces $B_{\textsc{pos}}$, {and} $\langle B_{\textsc{pos}}\rangle/\langle B_{\textsc{los}}\rangle$ $ = (\pi/4)/(1/2)$ $ = 1.58$ (cf. \citealt{pillai2016}).  This suggests that if the Zeeman measurements are treated as an accurate reference for $B_{\textsc{los}}$ ({cf.} \S~\ref{sec:zeeman}), then {the} DCF {measurements in Fig.~\ref{fig:compilation}} on average {overestimate} $B_{\textsc{pos}}$ by a factor $\sim 3-5$.  We note however that there is significant uncertainty on this value, and further that this is, if valid, a statistical correction to the ensemble of measurements {only.}   
{$\mathcal{M}_{A,{\rm median}}\sim 0.5$ would thus become} $\mathcal{M}_{A, {\rm median,corr}}\sim 1.5-2.5$, {suggesting} that turbulence is on average mildly super-Alfv\'enic to high densities. 
{However, if a conversion from 1D to 3D measures of both $B$ and $\sigma_{v,\textsc{nt}}$ is appropriate, these $\mathcal{M}_{A,{\rm median}}$ values will change by a factor of $\pi/4\sqrt{3}=0.45$, such that $\mathcal{M}_{A,{\rm median}}\sim 0.23$ and $\mathcal{M}_{A, {\rm median,corr}}\sim 0.7-1.3$.}

{{If turbulence were} indeed typically mildly super-Alfv\'enic, it is not clear why DCF
should {be accurate}.
(Of course, if the measurements shown in Fig.~\ref{fig:compilation} were entirely inaccurate, any inference drawn from them would be meaningless.)
An alternative interpretation of Fig.~\ref{fig:compilation}f is that some of the measurements are accurately measuring sub-Alfv\'enic turbulence, while others are {overestimates.}
We note that classical DCF is typically used only where $\sigma_{\theta} < 25^{\circ}$ \citep{ostriker2001}, which should select for stronger magnetic fields and more sub-Alfv\'enic turbulence.  {Examples exist of regions where DCF agrees well with other field strength measurements, notably in dense cores, including the massive core DR21(OH) \citep{crutcher1999,Hezarehetal2010,girart2013}, and the low-mass core sample of \citet{myers2021}. DCF also agrees with the modelled strength of an hourglass field in the hot molecular core G31.41+0.31 \citep{beltran2019}.}}  

Empirically, below densities $\sim 10^{7}$\,cm$^{-3}$, DCF appears, {at the level of calibration available between 2001 and 2021}, to provide an approximation to $B_{\textsc{pos}}$ which is likely to {on average be} overestimated by a factor of {a few.} 
The applicability of DCF at very high densities is particularly unclear, but may be better in better-resolved objects.  The DCF measurements made over the last 20 years broadly suggest that cloud turbulence is typically close to being Alfv\'enic ($\mathcal{M}_{A}\sim 0.5 - 2.5$), with no clear trend as a function of as density.
{A demonstration of if and where DCF is a somewhat accurate measure of magnetic field strength would provide fundamental information about $\mathcal{M}_{A}$ in molecular clouds.  Continuation of the current work to determine the accuracy and applicability of the method \citep{liu2021,liu2021a,skalidis2021a,li2021} is therefore essential.}

\subsubsection{\textbf{Magnetic field strength - density relationship}}
\label{sec:synth-Bvsn}

The relationship between magnetic fields and density ($B \propto n^{\kappa}$; eqs.~\ref{eq:B_vs_n}, \ref{eq:B_vs_n_emp}) is an important metric for determining whether ISM structures are magnetically supported {against collapse}.  Both the maximum value of $\kappa$ reached and the density $n_{0}$ at which $\kappa$ steepens are important diagnostics of the {magnetic criticality} of cloud substructures \citep[e.g.,][]{crutcher2012}.  \cite{crutcher2010} report a transition density of $n_{0}\,\sim\,300$\,cm$^{-3}$ and a power law slope of $\kappa\,=0.65\,\pm\,0.05$, based on a Bayesian analysis of both detections and non-detections of OH and CN Zeeman observations. This slope is consistent with self-similar collapse, suggesting that the magnetic field is unable to provide significant support against gravity.  
However, \cite{tritsis2015} argue that the uncertainties on the densities in the \cite{crutcher2010} sample are underestimated, that the models will be sensitive to the poorly constrained choice of $n_{0}$, and the assumption that $B$ is constant below $n_{0}$ is problematic. By relaxing these assumptions they obtain a $\kappa\sim 0.5$.  A further reanalysis has found $B_{0}=8.3\,\mu$G, $n_{0}=1125$\,cm$^{-3}$, and $\kappa = 0.72$ \citep{jiang2020}, and suggests that the observational uncertainties on both $B$ and $n$ prevent both $\kappa$ and $n_{0}$ from being accurately determined. 

\cite{li2015a} examined 100 dense clumps in both strong- and weak-field simulations, in both cases finding $\mu_{\Phi}\sim 2$ in, and a nearly spherical gravitational collapse of, the clumps.
\citet{li2015a}, \citet{mocz2017} and \citet{zhang2019} found that model clouds with both strong- and weak-field initial conditions produce $\kappa\sim 2/3$ at high gas densities, finding $n_{0}\sim 10^{4}$\,cm$^{-3}$, but that the transition may not be sharp, with $n_{0}$ not necessarily well-defined.  \citet{myers2021} suggest that the centrally-condensed spheroidal geometries of prestellar cores create $\kappa\approx 2/3$, despite their cores having relatively strong magnetic fields, further suggesting that the claim that $\kappa\approx 2/3$ indicates that clouds are in freefall collapse from very low densities should be treated with caution \citep[c.f.][]{tritsis2015}.

The \citet{crutcher2010} data set shows the $B-n$ relationship between an ensemble of clouds and cloud structures; the index $\kappa$ within individual structures may be quite different.  \citet{Lihuabai2015}, analysing dust polarisation maps (an indirect measure of magnetic field strength), find $\kappa = 0.4$ within the filamentary cloud NGC 6334 (see Fig.~\ref{fig:NGC6334}), and ascribe this to highly anisotropic cloud contraction along magnetic field lines.  \citet{wang2020a} find a similar result in the filamentary cloud IC 5146, $\kappa = 0.5^{+0.12}_{-0.13}$, while in simulations, \citet{zhang2019} find that within `affiliated structures', $\kappa=0.32\pm0.08$.  

{DCF results (Fig.~\ref{fig:compilation})} are broadly consistent with the \citet{crutcher2010} $B-n$ relationship. 
We see some suggestion that isolated cores have lower field strengths for a given density than do structures under {stellar} feedback, suggesting that 
$B_{0}$, $n_{0}$ and $\kappa$ may vary {with environment}.   {\citet{liu2021a} found a best-fit power law for their compilation of DCF results of $\kappa = 0.57\pm 0.03$, between the weak- and strong-field values.}

Most observational and theoretical works support an increase in $\kappa$ at high densities, associated with the transition to gravitational instability.  However, we cannot yet determine (i) the value of $n_{0}$, and if the transition is sharp, (ii) if $n_{0}$, $B_{0}$ and $\kappa$ vary between clouds or with environment, (iii) the nature of the physical relationship between $n_{0}$ and $n_{tr}$, the density of transition from cloud structures aligning preferentially field-parallel to field-perpendicular, (iv) differences between cloud-to-cloud variation in the $B-n$ relationship and evolution within clouds, (v) the extent to which the behavior of the $B-n$ relationship depends on the accretion history and collapse geometry of the cloud.

The best way to answer these questions is by obtaining more Zeeman measurements covering a wider range of densities. In particular the \citet{Thompson2019} OH absorption observations should span $100\,<\,n{_\mathrm{H}}\,<\,1000\,{\rm cm}^{-3}$, where $n_{0}$ is expected to lie. There are already indications that $\langle B_{\textsc{los}}\rangle$ for this sample is larger than in the CNM. More observations at high densities, particularly of more quiescent sources than the high mass star forming regions which make up the high density sightlines in the \cite{crutcher2010} sample, would also be helpful {(cf. \S~\ref{sec:forthcoming_instrumentation}).}
Finally, DCF 
{is} a comparatively inexpensive method of probing cloud fields, and can map entire star-forming regions (as opposed to individual pointings), and there are now many measurements available for statistical analyses.  If the biases and inherent uncertainties on DCF measurements can be understood, they may provide an independent data set to help constrain the $B-n$ relationship.

\subsubsection{\textbf{Correlations between cloud structure and magnetic field direction}}
\label{sec:synth-HROs}

As discussed in \S \ref{sec:hro} and \S \ref{sub:substructure}, most molecular clouds show a statistical preference for low column density ($N_{\mathrm H}$) filamentary substructure to align parallel to the magnetic field. In contrast, high column density structures preferentially align perpendicular to the field.  This change in alignment can be measured statistically using the Histograms of Relative Orientation (HROs) shape parameter ($\xi$ for comparisons with $\nabla N_{\mathrm{H}}$, $\zeta$ for $\nabla \rho$), or circular statistics such as the projected Rayleigh statistic $Z_{\mathrm{x}}$ (see \S \ref{sec:hro}).  
Generally, such a change in the relative orientation of the iso-$N_{\mathrm{H}}$ (2D projected) or iso-$n_{\mathrm{H}}$ (3D) contours is only seen in simulations with relatively strong magnetic fields ($\beta<1$) \citep{soler2013, seifried2020,kortgen_soler_2020}. It is, therefore, natural to ask what this transition can tell us about the energetic importance of the field.

\cite{soler2017b} examined the ideal MHD equations, studying the behavior of $\cos(\varphi)$, where $\varphi$ is the angle between $\nabla \rho$ and $\vec{B}$, and showed that both $\cos(\varphi) =\pm 1$ (parallel alignment) and $\cos(\varphi)\,=0$ (perpendicular alignment) represent equilibrium solutions. They found that the transition from parallel to perpendicular alignment occurs a) when $\nabla \cdot \vec{v}\,<\,0$, i.e.~in the presence of a convergent velocity field; b) where the large-scale magnetic field is strong enough to impose an anisotropic velocity field, i.e.~it imposes a preferred gas flow direction parallel to the field; and c) once the gas becomes strongly super-Alfv\'enic \citep{soler2017b, seifried2020}.

In many simulations, the relative orientation transition occurs where fields are strong on large scales (see Fig. \ref{fig:literature_hros}).  
For example, all three colliding flow simulations of \cite{chen2016}, which range from trans- to sub-Alfv\'enic and critical to sub-critical in the lower-density post-shock regions, show the transition.  \cite{seifried2020} found that only colliding flow simulations with initial field strengths $>$\,5$\mu$G showed a transition in alignment between $\nabla \rho$ and $\vec{B}$, while \cite{soler2013} found a transition in strong and intermediate magnetized ($\beta\,=\,0.1$ and 1) simulations, but not in the weakly magnetized simulation ($\beta\,=\,100$).  

\cite{chen2016} found that the transition density occurs when the gas becomes self-gravitating and gains enough kinetic energy to overcome the magnetic support ($\mu_{\Phi},\Malf>1$).  They suggest that by determining the density of the transition from parallel to perpendicular, one can estimate the density of equipartition between kinetic energy and gravity.  Turbulence models also show a tendency for gravity to aid perpendicular alignment both in sub- and super-Alfv\'{e}nic situations \citep{barretomota21}. 
{However, \citet{soler2017b} show that while the transition occurs only where gas motions are super-Alfv\'enic, $\mathcal{M}_{A}> 1$ is a necessary but not sufficient condition to trigger a change in alignment.} 

Larger-scale non-isothermal simulations that include feedback can show complicated alignment trends.  For the SILCC zoom-in simulations of \citet{seifried2020}, 4 out of 6 clouds show a clear transition in alignment between $\nabla \rho$ and $\vec{B}$. However, the cloud with the lowest mass-to-flux ratio, MC6 ($\mu_{\Phi,0}$\,=\,1.4), shows mostly no preferred orientation at high $n_{\mathrm H}$, while another cloud transitions to perpendicular alignment above $n_{\mathrm H} \sim 10^2$ cm$^{-3}$, then back to parallel alignment above $n_{\mathrm H} \sim 10^{3.3}$ cm$^{-3}$. \citet{IbanezMejia2021}, also in zoom-in simulations, find that the low-density gas is sub-Alfv\'{e}nic, with magnetic fields parallel to density structures. As the gas density increases, the gas becomes gradually super-Alfv\'{e}nic, and the angle between the magnetic field and density becomes random.

\begin{center}
    \centering
    \includegraphics[trim=0 20 0 15,width=0.5\textwidth]{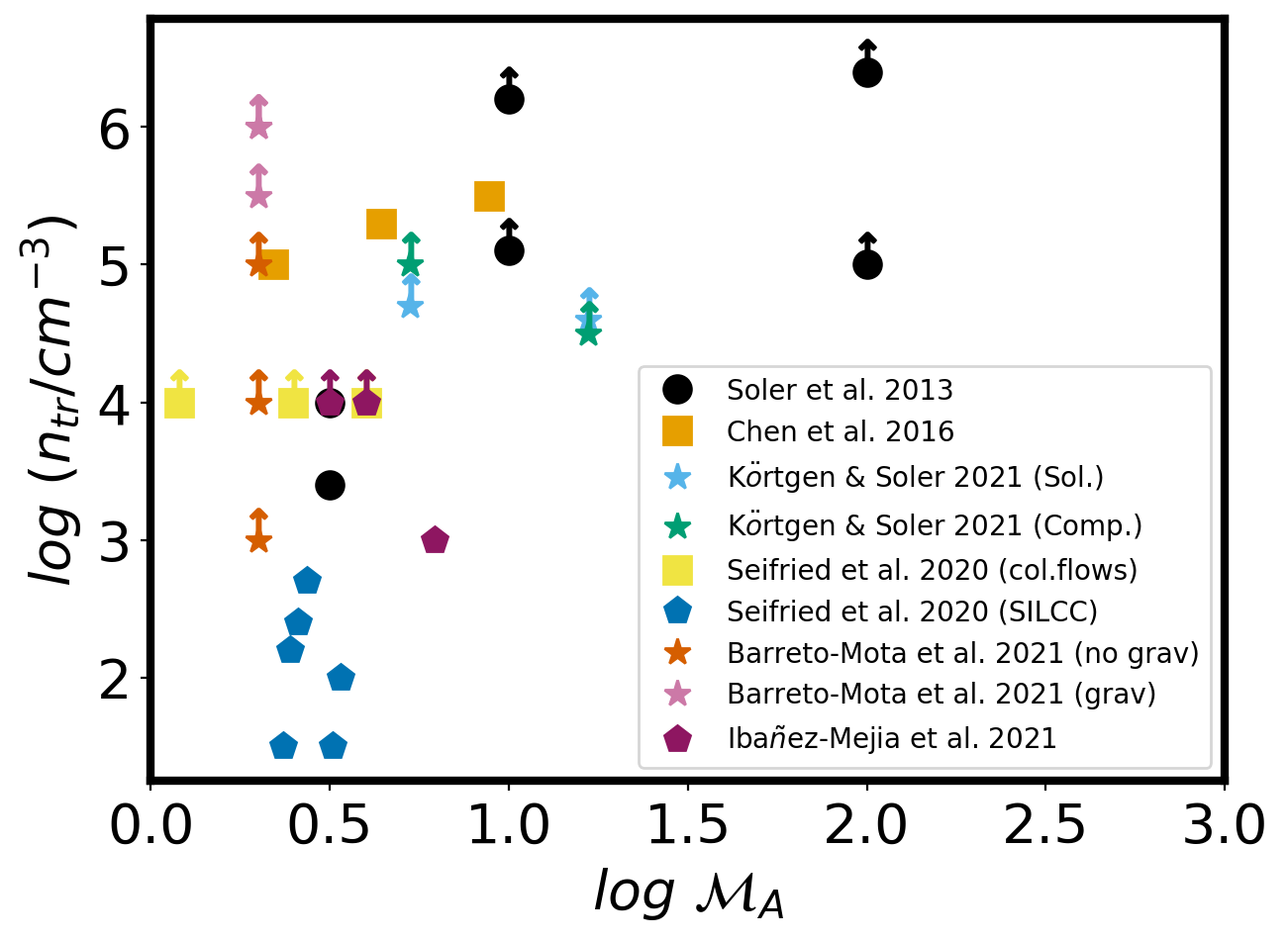}
    \captionof{figure}{Transition density, $n_{tr}$ as a function of Alfv\'{e}n Mach number $\mathcal{M}_{A}$ in various numerical works studying HROs. Symbols of the same shape represent similar setups: circles for decaying turbulence models, stars for driven turbulence, 
    {{squares}}
    for colliding flows, and 
    {{pentagons}}
    for zoom-in simulations. The symbols are accompanied by lower-limit arrows where no transition density was observed. In all cases, {$\mathcal{M}_{A}$ is that of the initial condition.}
    Exceptions are the {\citet{IbanezMejia2021} models}, where we roughly estimated $\mathcal{M}_{A}$ for n$>10^2$ from their Fig. 2. In the {\citet{seifried2020} models}, the mean $\mathcal{M}_{A}$ for n$>10^2$ ranges from 2.4--3.2. (D. Seifried, {priv. comm.}). }
    \label{fig:literature_hros}
\end{center}

In Fig. \ref{fig:literature_hros} we summarize the results of various numerical works that calculate HROs. The symbols indicate the transition density as a function of $\mathcal{M}_{A}$, with arrows showing lower limits where there was no observed transition.
{From this,} there seems to be no clear trend: different simulations with the same $\mathcal{M}_{A}$ reach very different results regarding the transition density. {Interpretation} of the results is further hindered by
{differing} definitions of $\mathcal{M}_{A}$ in the various models.
{Systematic} exploration of the parameter space {is necessary} to reach clear conclusions on the connection between the HRO and the energy balance of molecular clouds.

In summary, the presence of a transition in alignment from preferentially parallel to perpendicular implies the substructures formed in a cloud with a strong magnetic field ($\Malf \lesssim 1$) and with convergent velocity flows where the gas kinematics likely become super-Alfv\'enic. In self-gravitating molecular clouds, this convergent flow is likely related to the onset of gravitational collapse. At present
the 
{alignment transition column density cannot unambiguously be used} to estimate $\mathcal{M}_{A}$, due {both} to projection effects and
{to} the {complicated} relationship between energy balance and the cloud velocity field
not {being} fully understood.

% Figure 10 moved from here for arxiv version

Observations show that most clouds change in relative orientation from preferentially parallel to perpendicular or no preferred alignment at high $N_{\mathrm{H}}$~\citep{Planck_XXXV2016,soler2017}.  However, interpreting trends in individual clouds is challenging: the behavior of the relative orientation parameter $\xi$ or $Z_{\mathrm{x}}$ with $n_H$ shows considerable variation from cloud-to-cloud and also region-by-region within clouds (\citealt{soler2017,Soler2019}, see Fig. \ref{fig:HRO-obs}c).  Furthermore, the absence of a change in orientation with column density does not necessarily imply a weak super-Alfv\'enic field.  If the mean magnetic field direction is close to parallel to the line of sight, the field orientation will appear mostly random when projected on the plane-of-sky and show little correlation with the column density structure.  \cite{seifried2020} showed that the trends in relative orientation depend strongly on the viewing angle.

In the coming years, a fascinating study point will be whether there is a second transition from preferentially perpendicular alignment back to parallel alignment as the magnetic field lines are dragged in with the gas when a filament or core becomes highly supercritical, {necessarily occuring at much higher densities than $n_{tr}$, and on much smaller size scales}.  There are already hints of such a transition in some IRDCs and dense hub-like cluster forming regions  \citep{liu2018a,pillai2020}, similar to the strongly magnetized high-resolution {{simulations}} of a dense core by \cite{mocz2017}. {Of particular interest is whether there is a relationship between $n_{tr}$ and the size scale or density at which this second transition occurs.}

\subsection{3D fields and {cloud substructure} formation}

{Recent 3D observational and {numerical} studies of magnetic fields associated with molecular clouds provide us with hints regarding their formation and evolution mechanisms. 
{While all of the molecular cloud formation scenarios discussed in \S \ref{sub:formation}, gravitational instability of the Galactic disk, condensation from large-scale turbulence, colliding atomic flows, and shell expansion, may result in bending of magnetic field lines, the Parker instability and shell expansion/interaction models explicitly predict field line bending.}
In the shell expansion scenario, this bending happens for the {magnetic field} component perpendicular to the propagation direction of the shells, is at the location of formed filamentary structures, and helps to accumulate  
mass on these filamentary regions. The physical scale of this bending can be from a few pc (associated with small filaments) to tens of pc (associated with large filamentary molecular clouds). Depending on their orientation, these can result in observed {LOS} magnetic field reversal across these structures (from one side to the other along the short axis of the cloud). 

This {LOS} magnetic field reversal has recently been observed {using Faraday rotation} for three filamentary molecular clouds \citep{Tahanietal2018}, {and in Orion A} was previously observed with Zeeman measurements~\citep{heiles1989}. Magnetic field bending (a bow-shaped morphology) is believed to explain this reversal~\citep{heiles1989, Tahanietal2019}. Furthermore, magnetic field studies by \cite{Braccoetal2020} and {\citet{Tahanietal2022Pers}} or velocity observations by \cite{Bonneetal2020} support the predictions of expanding bubbles formation scenario. However, {statistical samples of such observations}
and more 
specific predictions from each of these models are required to help us better understand these formation scenarios and determine the most likely ones. {Such observations will be provided by} forthcoming RM {surveys:} the Polarization Sky Survey of the Universe's Magnetism~\citep[POSSUM;][]{Gaensler2010} 
with the Australian Square Kilometre Array Pathfinder (ASKAP), the Very Large Array Sky Survey (VLASS), or future Square Kilometre Array (SKA) observations~\citep{Healdetal2020}, with high source densities and improved uncertainties.  {Meanwhile, radiative transfer tools will allow testable predictions for both dust (e.g. DustPol \citealt{padovani2012}; Polaris, \citealt{reissl2016}) and line (e.g. Portal, \citealt{lankhaar2020}) emission polarization to be made for the cloud formation models.}}

{Within molecular clouds, observations indicate the presence of dynamically important magnetic fields that are strong enough to play an important role, influencing the orientation and characteristics of the cloud substructure \citep{Planck_XXXV2016,crutcher2010}. At low densities, elongated turbulent eddies, velocity shear, and the Lorentz force that inhibits motion perpendicular to the magnetic field lead to the formation of structures aligned parallel with the field, while convergent gas flows form high $N_{H}$ structures that preferentially align perpendicular to the cloud magnetic field. In both cases, strong magnetic fields set an energetically preferred direction of motion and of accretion onto dense structures {such as filaments, hubs or cores} (e.g. \citealt{Mocz2018}).}

{Understanding how the magnetic fields in dense cores connect to those in their parent filaments or hubs, and to cloud-scale fields, requires high-resolution, high-dynamic-range observations of entire clouds.   Accurate characterisation of 
{3D magnetic field geometries within dense cores}
is key to judging the applicability of 3D hourglass models \citep[e.g.][]{franzmann2017,myers2018}.  The amount of field curvature in collapsing cores should provide information on the level of flux-freezing \citep{basu2009}, and so on the magnetic initial conditions for fragmentation and the formation of hydrostatic cores.}

\subsection{Magnetic fields and star formation efficiency}
\label{sec:feedback}

Numerical models show that overall, magnetisation reduces the SFE {\citep[e.g.][]{KFreview19}}. On cloud scales, the magnetic field maintains filaments' coherence, and allows them to accrete along field lines. At the same time, it provides support against gravity, leading to a smaller number of cores than in an un-magnetized medium. However, on core scales and in the absence of feedback, models with different magnetic fields {convert core mass to stellar mass with similar efficiency}.

Nevertheless, the magnetic field can also affect the SFE indirectly by interacting with feedback. Protostellar outflows can mitigate turbulence dissipation \citep{baug2021} and can be significantly non-local \citep{offner2018}. They both decrease the SFE in the core from which they issue and reduce the efficiency of further core formation \citep{mathew2021}. In particular, \citet{offner2017} find that \emph{core-scale} SFE decreases, with efficiencies as low as 15\% (compared to $\sim$40--50\% in unmagnetized models; \citealt{offner2014}), and {lifetimes of Class 0 protostars} increase with magnetic field strength.

Larger-scale simulations suggest that the observed \emph{cloud-scale} star formation efficiencies {per free-fall time} of a few percent \citep[e.g.,][]{krumholz2007} can only be reproduced when turbulence, magnetic fields and jet/outflow feedback are all included \citep{wang2010,federrath2015}. Magnetic fields and outflow feedback in combination may reduce the SFE significantly more than does either effect alone \citep{wang2010, federrath2015}.

{A} correlation between the degree of magnetization and the cloud's star-formation efficiency may be {difficult} to confirm with {current} observations ({cf.} \S \ref{sec:sf_properties}). 
{However, ALMA observations of alternative measures of SFR, such as radio recombination lines or high-$J$ transitions of H$_3$CN \citep[e.g.][]{zhang2021}, may improve both SFR and SFE measurement and calibration over the next few years.}

\subsection{Questions for the next {decade}}

\subsubsection{\textbf{New and forthcoming instrumentation}}
\label{sec:forthcoming_instrumentation}

The next few years will provide a wealth of new instruments probing both dust continuum and spectral line polarisation.  Forthcoming single-dish emission polarimeters include
NIKA2-Pol {(1.15mm)} \citep{adam2018} at the Institut de Radioastronomie Millim\'etrique (IRAM) 30m Telescope, which is in the late stages of commissioning {\citep{ritacco2020,ajeddig2021}}; the A-MKID polarimeter (850$\mu$m, 350$\mu$m) on APEX \citep{otal2014}, which is currently being commissioned; TolTEC (2.1mm, 1.4mm, 1.1mm) on the Large Millimeter Telescope \citep[LMT;][]{bryan2018}, which {began} commissioning in {2021}; and Prime-Cam (850$\mu$m, 350$\mu$m) at the Fred Young Submillimeter Telescope (FYST)~\citep{CCAT2021}, under construction, with a target date of 2023/4. 
These instruments will have a few arcsecond to sub-arcminute resolution, and will map {up} to 60\% of the sky \citep{CCAT2021}.  There are also forthcoming extinction polarisation observations, {e.g.} the recent GPIPS data release \citep{Clemens2020}
and the upcoming PASIPHAE optical polarimetry survey
\citep{Mandarakasetal2019}.

ALMA can now
make CN Zeeman measurements, although only non-detections have been published \citep{vlemmings2019,harrison2021}.  The Five-hundred-meter Aperture Spherical Telescope (FAST) can measure the Zeeman effect in H\textsc{i} emission (probing the WNM/low-density CNM) and H\textsc{i} narrow self-absorption (HINSA; probing the intermediate-density CNM,
$\sim 10^{2}-10^{3}$ cm$^{-3}$; {\citealt{Ching2022}}).  In the longer term, the SKA will measure both the H\textsc{i} and OH Zeeman effects \citep{robishaw2015}, and the next-generation Very Large Array (ngVLA) will be able to observe the Zeeman effect in a range of transitions \citep{hull2018}.
ASKAP, the SKA, the Synthesis Telescope~\citep[ST;][]{Landeckeretal2019} at the Dominion Radio Astrophysical Observatory (DRAO), and the ngVLA will produce multi-wavelength RM catalogs with high source densities (20--30 deg$^{-2}$) and significantly reduced uncertainties: current LOFAR measurements \citep{VanEcketal2017} have uncertainties $\geq10\times$ smaller than those of \citet{Tayloretal2009}.
ALMA, the SKA and the ngVLA will also provide significant improvements in other spectropolarimetric measurements, particularly of the Goldreich-Kylafis effect. \citep[e.g.,][]{hull2018,stephens2020}.

{
The highest dynamic range in polarized dust emission can only be obtained with a large, cryogenically-cooled space telescope \citep[e.g.][]{andre2019a}.  Community support for polarimetric capabilities for any forthcoming FIR mission \citep{nasa_decadal} will be essential over the next few years.}

\subsubsection{\textbf{New and forthcoming simulations}}

While isolated cloud models remain valuable laboratories for studying core-scale effects, the recent approach of zooming onto molecular clouds from kpc-sized simulations has led to a deeper understanding of the complexity of the multi-phase, magnetized ISM. It has produced a clearer picture of the transitions between magnetically sub- and super-critical gas and probed the 3D morphology of the magnetic field around molecular clouds.

We expect to see this trend expand into galactic-scale MHD simulations in the coming years. Such models are abundant for un-magnetized galaxies (e.g., \citealt{Izquierdo2021, Renaud2013}) and show that molecular cloud properties depend on their location within the galaxy. Apart from the computational cost, the challenges in including magnetization will be i) understanding the role of the initial and boundary conditions for galactic magnetization and ii) correctly resolving the different dynamo mechanisms.

Significant new insights have arisen from including a cosmic ray fluid in MHD simulations {\citep[e.g.][]{girichidis2018a}}. Cosmic rays provide additional pressure to the galactic disk, affect the evolution of the Parker instability, and control $x_{e}$ in molecular clouds (hence also the coupling between the magnetic field and the gas). The importance of this addition to ISM models will likely soon produce a focused effort in this direction.

On smaller scales, considerable advances have been possible thanks to {models following the evolution of chemical abundances}. The challenges there lie in developing appropriate chemical networks for different environments while surpassing the associated computational burden. It will be exciting to see more efforts in this direction soon, especially towards coupling them with non-ideal MHD and radiation.

\subsubsection{\textbf{Key questions arising from this chapter}}

\textbf{What is the origin of cloud-scale magnetic fields, how do they connect to the larger-scale CNM, and how does their structure change over time?}
Understanding 3D field geometries in clouds will be key to answering this question.  New RM catalogs will enable the \citet{Tahanietal2018} technique to be further extended to more regions. This combined with forthcoming high-resolution, wide-area polarimetric mapping of molecular clouds 
will enable combined statistical analysis of LOS and POS magnetic field observations \citep[cf.][]{chen2018, Tahanietal2019} to determine the 3D field geometries in many molecular clouds. 
Gaia-estimated distances \citep{gaia2018} and large-scale velocity structure surveys such as GASKAP \citep{dickey2013} will further improve our understanding of large-scale field structure.  Forthcoming galactic-scale magnetic field simulations, as well as the inclusion of cosmic rays in models, will be crucial to understanding the origin and evolution of cloud-scale fields.  Improvements in radiative post-processing will lead to more testable predictions
{with} which to discriminate between models.

\textbf{Is there a critical density at which the ISM becomes gravitationally unstable, and how does it relate to the HRO transition density?}  Does this critical density vary between clouds, and does it depend on the cloud initial conditions?  Polarimetric mapping indicates that filaments and cores collapse anisotropically, with a preferential direction set by the magnetic field; how can this be reconciled with the self-similar $B\propto\rho^{2/3}$ result from Zeeman measurements?  The critical resource for understanding the $B-n$ relationship will be more Zeeman measurements in a range of environments, particularly around the transition density.  New single-dish instruments will produce systematic high-resolution, high-dynamic range dust polarisation observations of large samples of clouds, and so it will be possible to search for characteristic differences in behavior with gas density.  It will also be more possible to self-consistently analyse simulations over a wide range of scales, to determine how $\mathcal{M}_{A}$ and $\mu_{\Phi}$ vary with density.

\textbf{Is there observational evidence of a correlation between SFE and/or dense gas fraction and magnetic field properties, particularly $\mathcal{M_{A}}$ or $\mu_{\Phi}$?}  This question remains unanswered largely due to the significant difficulties of measuring both $\mathcal{M_{A}}$ and $\mu_{\Phi}$.  HRO analysis, while a promising tool, does not yet provide a simple diagnostic of either quantity.  DCF analysis must be better calibrated to provide a reliable measure of $\mathcal{M_{A}}$, and may perhaps only be trusted to measure $\mathcal{M_{A}}$ where it is independently verifiable that $\mathcal{M_{A}}< 1$, {although significant work on determining its form and calibration has recently been undertaken, and is continuing}.  Polarization fraction may provide a useful diagnostic for inclination angle and so a more accurate means of measuring $\mu_{\Phi}$.  However, we need better diagnostics, and more verification of our existing diagnostics against models, in order to properly exploit the forthcoming wealth of high-resolution observations of molecular clouds. 

\textbf{Do magnetic fields play a different role in high-mass star-forming cores than in low-mass star formation, and how important are non-ideal MHD effects?}
{While the idea of different initial conditions for low- and high-mass star formation is long-standing \citep[e.g.,][]{shu1987}, the detailed physics of high-mass star formation is only now becoming observable.}  ALMA observations are increasingly playing a crucial role in understanding the physics of high-mass star formation, and specifically the role played by the magnetic field.  This includes both linear and circular polarization measurements, and also the ability to simultaneously measure line and continuum emission in the same spectral set-up, allowing more rigorous comparison between observations and the expanding suite of astrochemical models.  Although explicit treatment of non-ideal MHD effects in large-scale simulations remains challenging, hybrid approaches involving non-ideal effects being introduced in zoom-in boxes allows the option of theoretically exploring the link between core-scale fields and the larger environment in which cores form.

\subsection{Summary}

Magnetic fields link gas on the largest scales in molecular clouds to that in gravitationally unstable star-forming cores.  It has become increasingly apparent that the traditional `magnetic fields versus turbulence' framing of star formation is a false dichotomy: recent results suggest that at low densities in molecular clouds, the Alfv\'en Mach number is not very different from 1, and so that magnetic fields and turbulent flows are not separable.
Magnetized turbulence dominates over gravity on large scales in molecular clouds, but on small scales gravity must win out.  There appears to be a magnetically sub- to super-critical break in cloud physics, but what gas density this occurs at, and if it is the same in all clouds, is not well-established.  There {also} appears to be a transition from cloud structures aligning preferentially parallel to the field, to aligning perpendicular to the field.  {Both effects appear to be associated with the transition to gravitational instability, and to within their large uncertainties appear to occur at similar densities,} but the precise nature of the relationship between these two transitions is not firmly established.

Magnetic fields are important in substructure formation, through both MHD turbulence and the direction of large-scale flows, and have a role in stabilising filaments and in directing accretion of material onto filaments and hubs, as well as providing direction to the collapse of cores.  Low-mass star formation may proceed from {prestellar} cores that are near magnetic criticality; there is limited evidence to suggest that high-mass star formation may proceed in a more supercritical fashion.  The interaction between magnetic fields and feedback is very important for setting star formation efficiency, and magnetic fields can couple outflow feedback over large scales in molecular clouds.

It is now apparent that magnetic fields are an integral part of the star formation process, and cannot be neglected even where they are playing an energetically subdominant role.  The wealth of forthcoming observations and simulations will over the next few years allow us to quantify our emerging understanding of how magnetic fields influence the outcome of the star formation process.

\FloatBarrier

\bigskip

\noindent\textbf{Acknowledgments} The authors would like to thank Che-Yu Chen, Daniel Seifried and Raphael Skalidis for helpful discussions, Qizhou Zhang for the SMA data shown in Fig.~\ref{fig:NGC6334}, and Richard Crutcher for the Zeeman measurements shown in Fig.~\ref{fig:compilation}.  K.P. is a Royal Society University Research Fellow, supported by grant number URF\textbackslash R1\textbackslash 211322. L.M.F.~is supported by a National Science and Engineering Research Council (NSERC) of Canada Discovery Grant. T.L. acknowledges the supports by National Natural Science Foundation of China (NSFC) through grants No.12073061 \& No.12122307, the international partnership program of Chinese Academy of Sciences through grant No.114231KYSB20200009, and Shanghai Pujiang Program 20PJ1415500. E.N. is supported by ERC ``Interstellar'' (Grant agreement 740120). 

%\bigskip

%\noindent\textbf{REFERENCES}
\bigskip
\parskip=0pt
{\small
\baselineskip=11pt

\bibliographystyle{pp7}

}
\end{document}